\documentclass[aps,prd,superscriptaddress,preprint,tightenlines,nofootinbib,floatfix]{revtex4}

\usepackage{graphicx} 
\usepackage{epsfig}
\usepackage{dcolumn}  
\usepackage{bm}       
\newcommand{\etal}{et al.}
\newcommand{\brat}{\ensuremath{{\mathcal B}_\mu / {\mathcal B}_e }}
\newcommand{\kpilndk}{\ensuremath{D^+ \rightarrow K^- \pi^+ \ell^+ \nu_{\ell} }}
\newcommand{\kpiendk}{\ensuremath{D^+ \rightarrow K^- \pi^+ e^+ \nu_{e} }}
\newcommand{\condBRe}{\ensuremath{\Gamma(\kpiendk)/\Gamma(D^+)}}
\newcommand{\condBRm}{\ensuremath{\Gamma(\kpimndk)/\Gamma(D^+)}}
\newcommand{\Hpls}{\ensuremath{H_+(\qsq)}}
\newcommand{\Hmin}{\ensuremath{H_-(\qsq)}}
\newcommand{\Hzer}{\ensuremath{H_0(\qsq)}}
\newcommand{\hzer}{\ensuremath{h_0(\qsq)}}
\newcommand{\HT}{\ensuremath{H_t(\qsq)}}
\newcommand{\Hspls}{\ensuremath{H^2_+(\qsq)}}
\newcommand{\Hsmin}{\ensuremath{H^2_-(\qsq)}}
\newcommand{\Hszer}{\ensuremath{H^2_0(\qsq)}}
\newcommand{\HsT}{\ensuremath{H^2_t(\qsq)}}
\newcommand{\hoHO}{\ensuremath{h_0(\qsq)H_0(\qsq)}}
\newcommand{\HTHO}{\ensuremath{H_0(\qsq)H_t(\qsq)}}

\newcommand{\hDFzer}{\ensuremath{h_0^{(d,f)}(\qsq)}}

\newcommand{\HDFint}{\ensuremath{\hDFzer\,\Hzer}}

\newcommand{\krz}{\ensuremath{K^{*0}}}
\newcommand{\krzb}{\ensuremath{\overline{K}^{*0}}}
\newcommand{\krzendk}{\ensuremath{D^+ \rightarrow \krzb e^+ \nu_e}}
\newcommand{\krzmndk}{\ensuremath{D^+ \rightarrow \krzb \mu^+ \nu_\mu}}
\newcommand{\krzlndk}{\ensuremath{D^+ \rightarrow \krzb \ell^+ \nu_\ell}}

\newcommand{\kpimndk}{\ensuremath{D^+ \rightarrow K^- \pi^+ \mu^+ \nu_{\mu} }}
\newcommand{\kpimundk}{\ensuremath{D^+ \rightarrow K^- \pi^+ \mu^+ \nu_{\mu} }}

\newcommand{\gevcsq}{\ensuremath{\textrm{GeV}/c^2}}

\newcommand{\gevcqd}{\ensuremath{\textrm{GeV}^2/c^4}}

\newcommand{\thv}{\ensuremath{\theta_\textrm{V}}}
\newcommand{\thl}{\ensuremath{\theta_\ell}}
\newcommand{\costhv}{\ensuremath{\cos\thv}}
\newcommand{\costhvsq}{\ensuremath{\cos^2\thv}}
\newcommand{\sinthv}{\ensuremath{\sin\thv}}
\newcommand{\costhl}{\ensuremath{\cos\thl}}

\newcommand{\sinthl}{\ensuremath{\sin\thl}}
\newcommand{\sinthlsq}{\ensuremath{\sin^2\thl}}
\newcommand{\qsq}{\ensuremath{q^2}}

\newcommand{\bw}{\ensuremath{\beta}}
\newcommand{\mkpi}{\ensuremath{m_{K\pi}}}

\newcommand{\rtwo}{\ensuremath{r_2}}
\newcommand{\rthree}{\ensuremath{r_3}}
\newcommand{\rvee}{\ensuremath{r_\textrm{v}}}
\newcommand{\mysection}[1]{\section{#1}}
\newcounter{saveeqn}

\begin{document}

%\preprint line(s) will be ignored for PRL/PRD
%\preprint{CLEO Draft YY-NNA} % For paper draft CBX YY-NN -> Draft YY-NNA
%\preprint{CLEO CONF YY-NN}   % For conference papers
%\preprint{ICHEP ABSnnn}      % For conference papers

\preprint{CLNS 10/2063}       % for CLNS notes
\preprint{CLEO 10-01}         % for CLNS notes

\title{Analysis of  {\boldmath{\kpiendk{}}} and {\boldmath{\kpimndk{}}} Semileptonic Decays}

\author{R.~A.~Briere}
\author{H.~Vogel}
\affiliation{Carnegie Mellon University, Pittsburgh, Pennsylvania 15213, USA}
\author{P.~U.~E.~Onyisi}
\author{J.~L.~Rosner}
\affiliation{University of Chicago, Chicago, Illinois 60637, USA}
\author{J.~P.~Alexander}
\author{D.~G.~Cassel}
\author{S.~Das}
\author{R.~Ehrlich}
\author{L.~Fields}
\author{L.~Gibbons}
\author{S.~W.~Gray}
\author{D.~L.~Hartill}
\author{B.~K.~Heltsley}
\author{J.~M.~Hunt}
\author{D.~L.~Kreinick}
\author{V.~E.~Kuznetsov}
\author{J.~Ledoux}
\author{J.~R.~Patterson}
\author{D.~Peterson}
\author{D.~Riley}
\author{A.~Ryd}
\author{A.~J.~Sadoff}
\author{X.~Shi}
\author{W.~M.~Sun}
\affiliation{Cornell University, Ithaca, New York 14853, USA}
\author{J.~Yelton}
\affiliation{University of Florida, Gainesville, Florida 32611, USA}
\author{P.~Rubin}
\affiliation{George Mason University, Fairfax, Virginia 22030, USA}
\author{N.~Lowrey}
\author{S.~Mehrabyan}
\author{M.~Selen}
\author{J.~Wiss}
\affiliation{University of Illinois, Urbana-Champaign, Illinois 61801, USA}
\author{M.~Kornicer}
\author{R.~E.~Mitchell}
\author{M.~R.~Shepherd}
\author{C.~M.~Tarbert}
\affiliation{Indiana University, Bloomington, Indiana 47405, USA }
\author{D.~Besson}
\affiliation{University of Kansas, Lawrence, Kansas 66045, USA}
\author{T.~K.~Pedlar}
\author{J.~Xavier}
\affiliation{Luther College, Decorah, Iowa 52101, USA}
\author{D.~Cronin-Hennessy}
\author{J.~Hietala}
\author{P.~Zweber}
\affiliation{University of Minnesota, Minneapolis, Minnesota 55455, USA}
\author{S.~Dobbs}
\author{Z.~Metreveli}
\author{K.~K.~Seth}
\author{X.~Ting}
\author{A.~Tomaradze}
\affiliation{Northwestern University, Evanston, Illinois 60208, USA}
\author{S.~Brisbane}
\author{J.~Libby}
\author{L.~Martin}
\author{A.~Powell}
\author{P.~Spradlin}
\author{G.~Wilkinson}
\affiliation{University of Oxford, Oxford OX1 3RH, UK}
\author{H.~Mendez}
\affiliation{University of Puerto Rico, Mayaguez, Puerto Rico 00681}
\author{J.~Y.~Ge}
\author{D.~H.~Miller}
\author{I.~P.~J.~Shipsey}
\author{B.~Xin}
\affiliation{Purdue University, West Lafayette, Indiana 47907, USA}
\author{G.~S.~Adams}
\author{D.~Hu}
\author{B.~Moziak}
\author{J.~Napolitano}
\affiliation{Rensselaer Polytechnic Institute, Troy, New York 12180, USA}
\author{K.~M.~Ecklund}
\affiliation{Rice University, Houston, Texas 77005, USA}
\author{J.~Insler}
\author{H.~Muramatsu}
\author{C.~S.~Park}
\author{E.~H.~Thorndike}
\author{F.~Yang}
\affiliation{University of Rochester, Rochester, New York 14627, USA}
\author{S.~Ricciardi}
\affiliation{STFC Rutherford Appleton Laboratory, Chilton, Didcot, Oxfordshire, OX11 0QX, UK}
\author{C.~Thomas}
\affiliation{University of Oxford, Oxford OX1 3RH, UK}
\affiliation{STFC Rutherford Appleton Laboratory, Chilton, Didcot, Oxfordshire, OX11 0QX, UK}
\author{M.~Artuso}
\author{S.~Blusk}
\author{S.~Khalil}
\author{R.~Mountain}
\author{T.~Skwarnicki}
\author{S.~Stone}
\author{J.~C.~Wang}
\author{L.~M.~Zhang}
\affiliation{Syracuse University, Syracuse, New York 13244, USA}
\author{G.~Bonvicini}
\author{D.~Cinabro}
\author{A.~Lincoln}
\author{M.~J.~Smith}
\author{P.~Zhou}
\author{J.~Zhu}
\affiliation{Wayne State University, Detroit, Michigan 48202, USA}
\author{P.~Naik}
\author{J.~Rademacker}
\affiliation{University of Bristol, Bristol BS8 1TL, UK}
\author{D.~M.~Asner}
\author{K.~W.~Edwards}
\author{J.~Reed}
\author{K.~Randrianarivony}
\author{A.~N.~Robichaud}
\author{G.~Tatishvili}
\author{E.~J.~White}
\affiliation{Carleton University, Ottawa, Ontario, Canada K1S 5B6}
\collaboration{CLEO Collaboration}
\noaffiliation

%\date{\today}
\date{April~22,~2010}

\begin{abstract}
Using a large sample ($\approx$ 11800 events) of \kpiendk{} and \kpimndk{} 
decays collected by the CLEO-c detector running at the $\psi(3770)$, we
measure the helicity basis form factors free
from the assumptions of spectroscopic pole dominance and provide new, accurate 
measurements of the absolute branching fractions for \krzendk{} and \krzmndk{} 
decays. We find branching fractions which are consistent with previous world 
averages. Our measured helicity basis form factors are consistent with the 
spectroscopic pole dominance predictions for the three main helicity basis
form factors describing \krzlndk{} decay. The ability to 
analyze \kpimndk{} allows us to make the first non-parametric measurements of 
the mass-suppressed form factor. Our result is inconsistent with existing 
Lattice QCD calculations. 
Finally, we measure the form factor that controls non-resonant $s$-wave 
interference with the \krzlndk{} amplitude and search for evidence of 
possible additional non-resonant $d$- or $f$-wave interference with the \krzb.
\end{abstract}

\pacs{13.20.Fc, 12.38.Qk, 14.40.Lb}
\maketitle

\mysection{\label{intro}INTRODUCTION}
We present new measurements of the \krzendk{} and \krzmndk{} absolute branching
fractions, their ratio, and measurements of the semileptonic form 
factors controlling these decays.\footnote{Throughout this paper the 
charge conjugate is implied when a decay mode of
a specific charge is stated.}$^,$\footnote{We reconstruct \krzlndk{}
modes as \kpilndk{} decays, and use the
Clebsch-Gordan factor 1.5 to correct for $\krzb \rightarrow \overline{K}^{0} \pi^0$ decays, which we
do not detect.}
Exclusive charm semileptonic decays provide particularly simple tests of over
decay dynamics since long distance effects only enter through the hadronic form
factors~\cite{bigi}. A wide variety of theoretical methods have
been brought to bear on the calculation of these form factors including quark
models~\cite{quark}, QCD sum rules~\cite{ball}, Lattice QCD~\cite{lqcd}, analyticity~\cite{analyticity},
and others~\cite{slovenia}.
Using a technique developed by FOCUS~\cite{helicity-focus}, we present
non-parametric measurements of the \qsq{} dependence of the
helicity basis form factors that give an amplitude for the $K^- \pi^+$
system to be in any one of its possible angular momentum states where \qsq{} is
the invariant mass squared of the lepton pair in the decay.
The ultimate goal of this study is to obtain a better understanding
of the semileptonic decay intensity.  
  
CLEO-c produces $D$ mesons at the  $\psi$(3770), which ensures
a pure $D \overline{D}$ final state with no additional final state hadrons. 
In events where the \kpilndk{} is produced against a fully reconstructed $D^-$
the missing neutrino can be reconstructed with unparalleled precision using
energy-momentum balance. Hence, CLEO-c data offer unparalleled \qsq{} and 
decay angle resolution allowing one
to resolve fine details in the structure of these form factors without
the complications of a deconvolution procedure. The various helicity basis form
factors are distinguished based on their contributions to the decay angular 
distribution.

\begin{figure}[htbp]
\begin{center}
\includegraphics[width=3.2in]{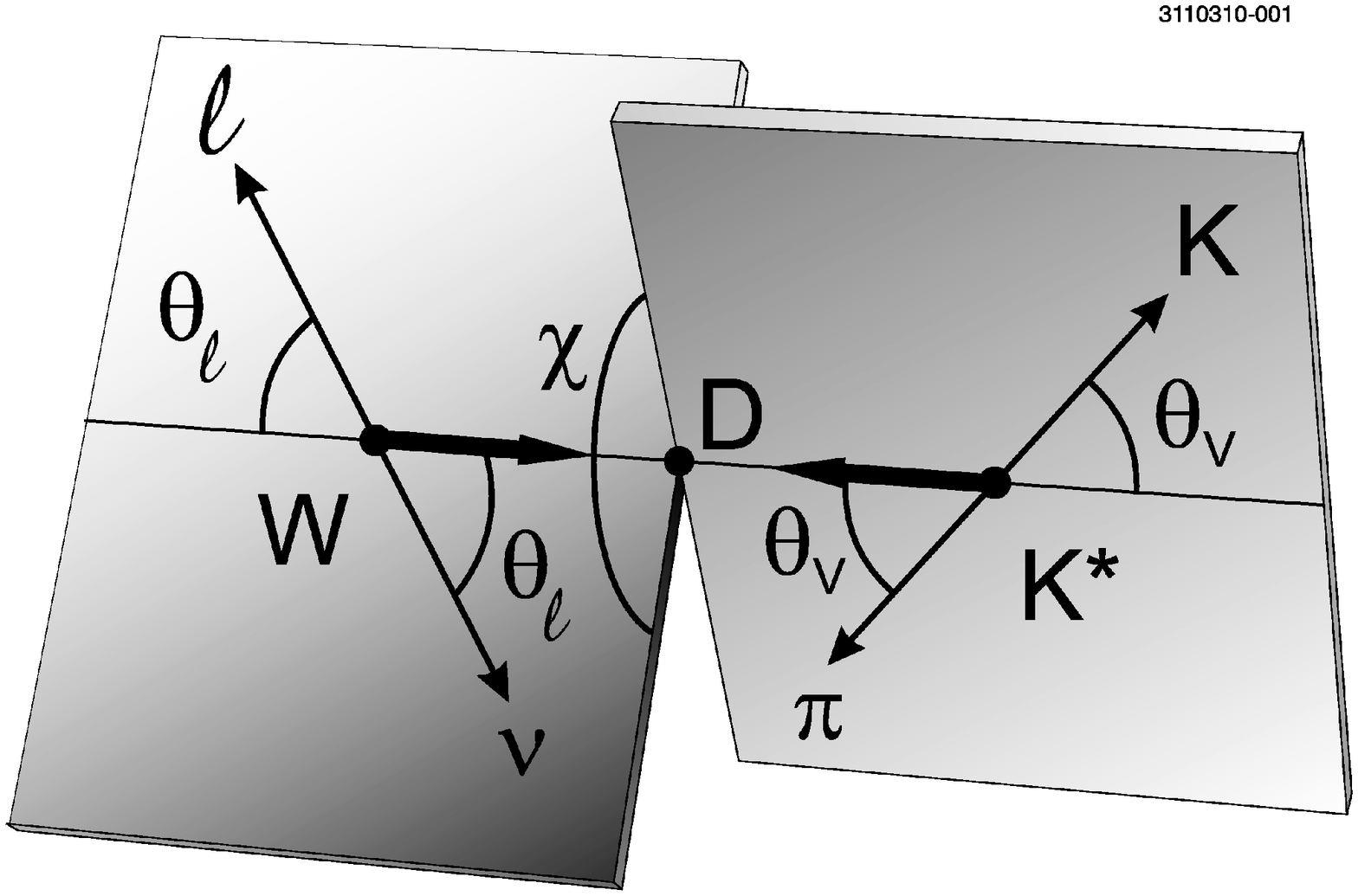}
\caption{Definition of the ${ \theta_{V}}$, ${ \theta_{\ell}}$, 
and $\chi$ angles.}
\label{fig:kinem-definition}
\end{center}
\end{figure}

The amplitude $\mathcal A$ for the semileptonic decay \kpilndk{} is described
by five kinematic quantities: 
\qsq{}; the kaon-pion mass (\mkpi{}); the kaon helicity angle ($\thv$), 
which is computed as the angle between the $\pi$ and the $D$ direction in 
the $K^- \pi^+$ rest frame;
the lepton helicity angle ($\thl$), which is computed as the angle between 
the $\nu_{\ell}$ and the $D$ direction in the $\ell^+ \nu_e$
rest frame;  
and the acoplanarity angle between the two decay planes ($\chi$). The
decay angles are illustrated in Fig.~\ref{fig:kinem-definition}.
The amplitude $\mathcal A$ can be expressed in terms of four helicity
amplitudes representing the transition to the vector \krzb{}:
 \Hpls{}, \Hmin{}, \Hzer{}, \HT{} and a fifth form factor, \hoHO{} describing
a non-resonant,~$s$-wave \kpilndk{} contribution. 

The diferential decay width for the 4-body semileptonic process is
\begin{equation}
{d^5 \Gamma \over d \costhl ~d\costhv ~d \chi~ d\qsq~d\mkpi^2}={|\mathcal A|^2 K P_{\ell} P^* \over 256 \pi^6 m_D^2 \sqrt{q^2}\mkpi}\,,
\label{Gamma}
\end{equation}
where $|\mathcal A|^2$ is the decay intensity, $K$ is the $K^- \pi^+$ momentum in the $D^+$ rest frame, $P^*$ is the momentum
of the kaon in the $K^- \pi^+$ rest frame,
and $|\vec P_\ell|$ is the 
momentum of the $\ell^+$ in the $\ell^+ \nu$ rest frame. 
Upon integration over
$\chi$, the differential decay width is proportional to: 
\begin{widetext}
\begin{eqnarray}
   \int { |\mathcal{A}|^2 d\chi }
      = \frac{q^2 - m_\ell ^2}{8} \left\{ \begin{array}{l}
              ((1 + \costhl) \sinthv)^2 |H_+(q^2)|^2 |\bw|^2  \\ 
            + ((1 - \costhl) \sinthv)^2 |H_-(q^2)|^2 |\bw|^2  \\ 
            + (2 \sinthl \costhv)^2 |H_0(q^2)|^2 |\bw|^2  \\ 
            + 8~\underline{\sinthlsq \costhv H_0(q^2)h_0(q^2)
                {\mathop{\rm Re}\nolimits}\{Ae^{-i\delta} \bw \}}
            \end{array} \right\} \nonumber \\
  \mbox{} \label{amp1} \\
  \mbox{} +\frac{|\bw|^2}{8}(q^2 - m_\ell^2)\frac{m_\ell^2}{q^2}
                                 \left\{ \begin{array}{l}
              (\sinthl \sinthv)^2 |H_+(q^2)|^2   
                + (\sinthl \sinthv)^2 |H_-(q^2)|^2  \\ 
            + (2 \costhl \costhv)^2 |H_0(q^2)|^2  \\ 
            + (2 \costhv)^2 |H_t(q^2)|^2
            + 8\costhl \costhvsq H_0(q^2) H_t(q^2)
            \end{array} \right\} \, \nonumber
\end{eqnarray}
\end{widetext}

The \HT{} form factor, 
which appears in the second term of Eq.~(\ref{amp1}),
is helicity suppressed by a factor of $m_\ell^2/q^2$. 
The mass-suppressed terms are negligible for \kpiendk{} but can be measured 
in \kpimndk{}. The \HT{} form factor can only be effectively measured 
in \kpimndk{} decays at low \qsq{}
where the mass suppression effects are least severe. 
The semimuonic to semielectric 
branching ratio is sensitive to the magnitude of the \HT{} form factor.

We study the form factor of the non-resonant, spin
zero, $s$-wave component to \krzmndk{} first described in Ref.~\cite{anomaly}. 
According to the model of
Ref.~\cite{formfactor},  %$2.4 \pm 0.7\,\%$ 
2.4\% of the decays in
the mass range $0.8~\gevcsq{} < \mkpi{} < 1.0~\gevcsq{}$ are due to
this $s$-wave component\cite{absBeCLEO}, where \mkpi{} is the $K^- \pi^+$ mass. 
The underlined term in Eq.~(\ref{amp1}) represents the interference between
the $s$-wave, $K^- \pi^+$ amplitude and the \krzb{} amplitude, represented
as a simplified, Breit-Wigner function of the form:
\begin{equation}
\bw = \frac{\sqrt{m_0 \Gamma} \left(\frac{P^*}{P_0^*}\right)}
	   {m_{K\pi}^2 - m_0^2 + i m_0 \Gamma 
	      \left(\frac{P^*}{P_0^*}\right)^3}\, 
\label{BW-beta}
\end{equation}
where $P^*$ is the kaon momentum in the $K^- \pi^+$ rest frame, and
$P_0^*$ is the value of $P^*$ when the $K^- \pi^+$ mass is equal to the 
\krzb{} mass\footnote{We are using a $p$-wave Breit-Wigner form with a
width proportional to the cube of the kaon momentum in the kaon-pion
rest frame.  Our Breit-Wigner intensity is  proportional to $P^{*3}$ as 
expected for a $p$-wave Breit-Wigner resonance. Two powers of $P^*$ come explicitly from the $P^*$ in
the numerator of the amplitude and one power arises from the 4-body
phase space as shown in Eq.~(\ref{Gamma}). We are not including additional, small corrections such as the Blatt-Weisskopf barrier penetration factor.}.
 
The $s$-wave form factor is denoted as \hzer{} in the underlined piece of Eq.~(\ref{amp1}).
Following Ref.~\cite{anomaly} we model the $s$-wave contribution as an amplitude
with a phase ($\delta$) and modulus ($A$) that are independent of \mkpi{}.
We have dropped the second-order, $s$-wave intensity contribution ($\propto |A|^2$) in Eq.~(\ref{amp1}) since
$A \ll |\bw|$.

The $\chi$ integration significantly simplifies the intensity by eliminating
all interference terms between different helicity states of the virtual $W^+$
with relatively little loss in form factor information. 

The four helicity basis form factors for the \krzmndk{} component are
generally written~\cite{KS} as linear combinations of a vector ($V(\qsq{})$)
and three axial-vector ($A_{1,2,3}(\qsq{})$) form factors 
according to 

\begin{widetext}
\begin{eqnarray}
H_\pm(\qsq) &=&
   (M_D+\mkpi)A_1(\qsq)\mp 2{M_D K\over M_D+m_{K\pi}}V(\qsq) \,,
                               \nonumber \\
H_0(\qsq) &=&
   {1\over 2\mkpi\sqrt{\qsq}}
   \left[
    (M^2_D -m^2_{K\pi}-\qsq)(M_D+\mkpi)A_1(\qsq) \frac{}{} \right. \nonumber \\
          & & \mbox{} \hspace{2cm} \left.
    -4{M^2_D K^2\over M_D+\mkpi}A_2(\qsq) \right] \,,\label{helicity} \\ 
H_t(\qsq) &=&
   {M_D K\over m_{K\pi}\sqrt{\qsq}}
   \left[ (M_D+m_{K\pi})A_1(\qsq)
    -{(M^2_D -m^2_{K\pi}+\qsq) \over M_D+m_{K\pi}}A_2(\qsq) \right. \nonumber\\
          & & \mbox{} \hspace{2cm} \left.
    +{2\qsq\over M_D+m_{K\pi}}A_3(\qsq) \right] \,,\nonumber 
\end{eqnarray}
\end{widetext}
where $M_D$ is the mass of the $D^+$ and $K$ is the momentum of the $K^- \pi^+$
system in the rest frame of the $D^+$.
In the Spectroscopic Pole Dominance (SPD) model \cite{KS,formfactor}, these
axial and vector form factors are given by 
\begin{eqnarray}
V\left( {q^2 } \right) = \frac{{V(0)}}{{1 - q^2 /M_V^2 }}{\rm{ ~~;~~  }}A_{{\rm{1,2,3}}} \left( {q^2 } \right) = \frac{{A_{{\rm{1,2,3}}} (0)}}{{1 - q^2 /M_A^2 }}~~,~~ \label{SPDpole}
\end{eqnarray}
where
$M_V = 2.1~\gevcsq$ and $M_A = 2.5~\gevcsq$.  The SPD model allows one to parameterize the 
\Hmin{}, \Hpls{}, \Hzer{}, and \HT{} form factors using just three 
parameters, which are ratios of form factors taken at 
$\qsq = 0$\,: $\rvee \equiv V(0)/ A_1(0),\ \rtwo \equiv A_2(0)/A_1(0)$ and
$\rthree{} = A_3(0)/A_1(0)$. There are accurate measurements~\cite{formfactor} 
of \rvee{} and \rtwo{}, but very little is known about \rthree{}, which is an 
important motivation for this work. 
 
In this paper, we use a \emph{projective weighting}
technique~\cite{helicity-focus} to disentangle and directly measure the \qsq{}
dependence of these helicity basis form factors free from parameterization.
We provide information on the six form factor products
$H_\pm^2(\qsq)$, \Hszer{}, \hoHO{}, \HsT{} and \HTHO{} in bins 
of \qsq{} by projecting out the associated angular factors given
by Eq.~(\ref{amp1}).
We next describe some of the experimental and analysis details used for these
measurements.

\mysection{\label{exp} EXPERIMENTAL AND ANALYSIS DETAILS}
The CLEO-c detector~\cite{detector} consists of a six-layer inner
stereo-wire drift chamber,
a 47-layer central drift chamber, a ring-imaging Cerenkov detector
(RICH), and a cesium iodide electromagnetic calorimeter inside 
a superconducting
solenoidal magnet providing a 1.0~T magnetic field. The tracking chambers
and the electromagnetic calorimeter cover 93\% of the full solid angle.
The solid angle coverage for the RICH detector is 80\% of $4\pi$. 
Identification
of the charged pions and kaons is based on measurements of specific
ionization ($dE/dx$) in the main drift chamber and RICH information.
Electrons are identified using the ratio of the energy deposited in the electromagnetic calorimeter to
the measured track momentum ($E/p$) as well as  $dE/dx$ and RICH information.
Although there is a muon
detector in CLEO, it was optimized for b-meson semileptonic decay,
and is ineffectual for charm semileptonic decay since a muon from charm 
particle decay will typically range out in the first layer of iron in the muon shield.

In this paper, we use 818~pb$^{-1}$ of data taken at the $\psi(3770)$
center-of-mass energy with the CLEO-c detector at the Cornell Electron Storage
Ring (CESR) $e^+ e^-$ collider, which corresponds to a (produced) sample of
1.8 million $D^+D^-$ pair events~\cite{HadronicBrFraction}. 

We select the events containing a $D^-$ decaying into
 one of the following six decay modes:
$D^- \rightarrow K^0_S\pi^-$, $D^- \rightarrow K^+\pi^-\pi^-$,
$D^- \rightarrow K^0_S\pi^-\pi^0$, $D^- \rightarrow K^+\pi^-\pi^-\pi^0$,
$D^- \rightarrow K^0_S\pi^-\pi^-\pi^+$, and $D^- \rightarrow K^-K^+\pi^-$ 
along with a 4-body semileptonic candidate.
To avoid  complications due to
having two or more \kpilndk{} decay candidates in the event, we select
the decay candidate with the smallest $ |M_\textrm{bc}-M_{D^{-}}|$ value where $M_\textrm{bc}$
is the beam-constrained mass. 
The beam-constrained mass $M_\textrm{bc}$ is defined as 
$M_\textrm{bc}c^2 = \sqrt{|(E_\textrm{beam})^2 - c^2 P^2_\textrm{D}|}$
where $E_\textrm{beam}$ is the beam energy and $P_D$ is the D-tag momentum. 
More details on selecting the tagging $D^-$ candidates as well as identifying
$\pi^0$ and $K^0_S$ candidates are described in Ref.~\cite{HadronicBrFraction}.

We used extensive
Monte Carlo (MC) studies to design efficient, 
background-suppressing selections. 
The \kpilndk{} reconstruction starts by requiring three well-measured tracks
not associated with the  tagging $D^-$ decay. 
In order to select semileptonic decays, we require a minimal missing momentum
and energy of 50~MeV/$c$ and 50~MeV, respectively. Both the minimal 
missing momentum and energy are 
calculated using the center-of-mass momentum and energy.
In order to reduce backgrounds from charm decays with missing $\pi^0$~'s,  
we require an unassociated shower energy of less than 250~MeV. The 
unassociated shower energy refers
to electromagnetic showers, which are statistically separated from all 
measured, charged tracks. 
Charged kaons and pions are required to have momenta of at least 50~MeV/$c$ 
and are
identified using $dE/dx$ and RICH information. We require that the pion 
deposits a 
shower energy, which is inconsistent with the electron hypothesis. 

Electron candidates are
required to have momenta of at least 200~MeV/$c$, lie in the good shower containment 
region ($|\cos \theta| < 0.9$), and pass a requirement on a 
likelihood variable that combines $E/p$, $dE/dx$, and RICH information. Our simulations indicate
that contamination of our kaon sample due to pions is less than 0.06\% using this likelihood variable. 
The only final state particle not detected is
the neutrino in the semileptonic decay. The neutrino four-momentum vector
can be reconstructed from the missing energy and momentum in the event.
The \qsq{} resolution, predicted by our Monte Carlo simulation, is roughly Gaussian with an r.m.s.
width of 0.02 \gevcqd{}, which is negligible on the scale that we will bin our data.

For \kpimndk ~candidates, it is difficult to distinguish the $\pi^+$  track from the
$\mu^+$ track. Because \kpimndk{} decay is strongly dominated by $\krzb \rightarrow K^- \pi^+$,  
which is a relatively narrow resonance, we select the positive track with the 
smallest $|\mkpi{} - m_{\krzb}|$ as the pion and the other track as the muon. Our Monte Carlo
studies concluded that this \krzb{} arbitration approach was correct 84\% of the time and works 
better than pion-muon discrimination based on the electromagnetic calorimeter response. 

\begin{figure}[htbp]
\includegraphics[width=2.8in]{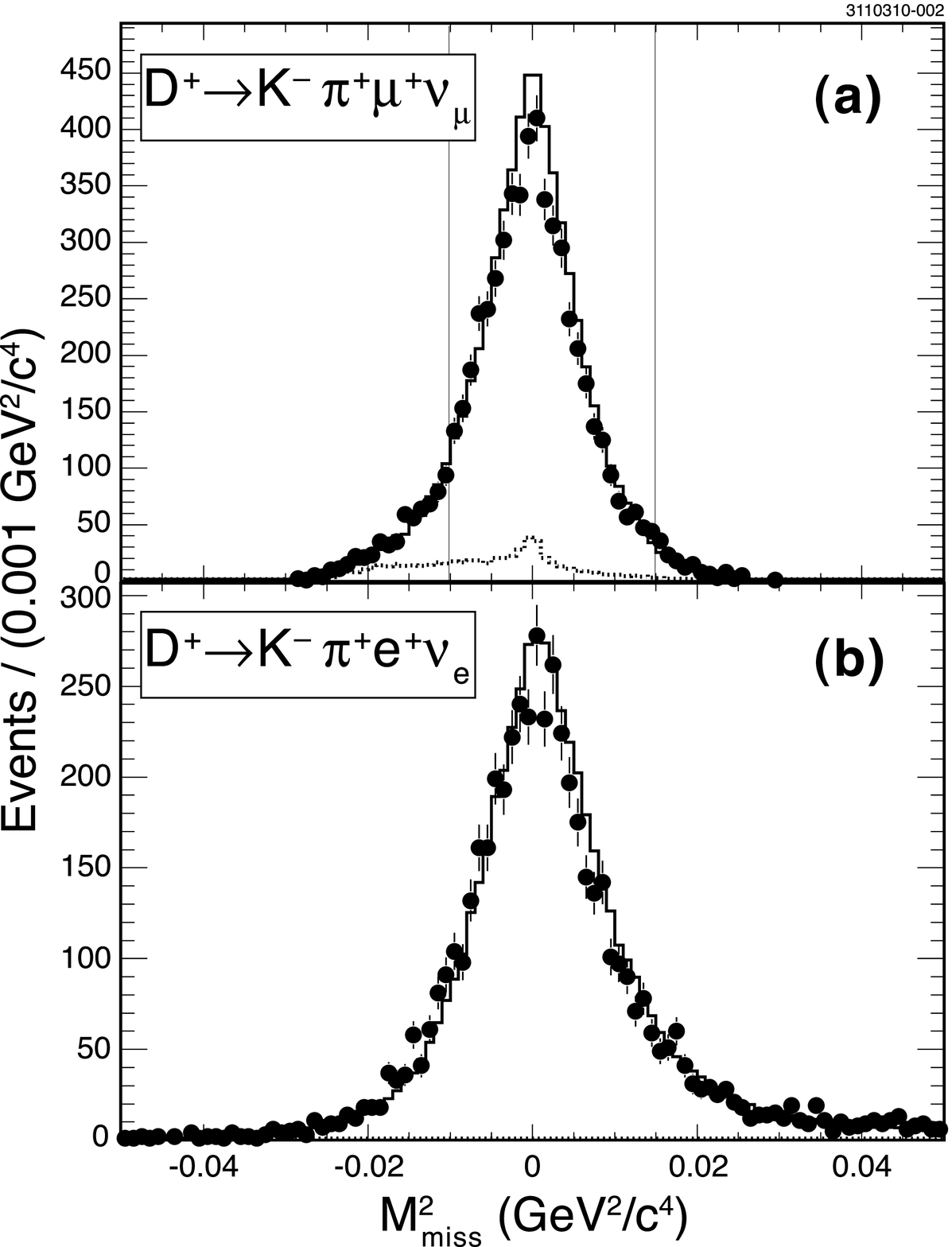}
\caption{ The $M^2_\textrm{miss}$ distributions for events satisfying our \emph{nominal}
\kpilndk{} selection requirements apart from the $M^2_\textrm{miss}$ requirement. (a) 
shows the $M^2_\textrm{miss}$ distribution for \kpimndk{} candidates, while 
(b) shows the $M^2_\textrm{miss}$ distribution for \kpiendk{} candidates.
For \kpimndk{} candidates, we require that
$M^2_\textrm{miss}$ lies between the vertical lines. This cut is placed asymmetrically
on our semimuonic sample to suppress cross-feed from \kpiendk{}. In each plot, 
the solid histogram shows the signal plus 
background distribution predicted 
by our Monte Carlo simulation, while the dashed histogram shows the 
predicted background component.
   \label{mmsq}}
\end{figure}

We apply a variety of additional requirements to suppress backgrounds in \kpimndk{} candidates.
We require that the muon is inconsistent with the electron hypothesis according to the electron 
likelihood variable. We require that missing momentum ($P_\textrm{miss}$) lies within 20~MeV of 
the missing energy ($E_\textrm{miss}$). For \kpimndk{} candidates, we also require  
$-0.01 <  M^2_\textrm{miss} < 0.015~\textrm{GeV}^2/c^4$. The $M^2_\textrm{miss}$ distributions for muon and 
$-0.01 <  M^2_\textrm{miss} < 0.015~\textrm{GeV}^2/c^4$. The $M^2_\textrm{miss}$ distributions for muon and 
electron candidates are illustrated in Fig.~\ref{mmsq}.

\begin{figure}[htbp]
\includegraphics[width=2.8in]{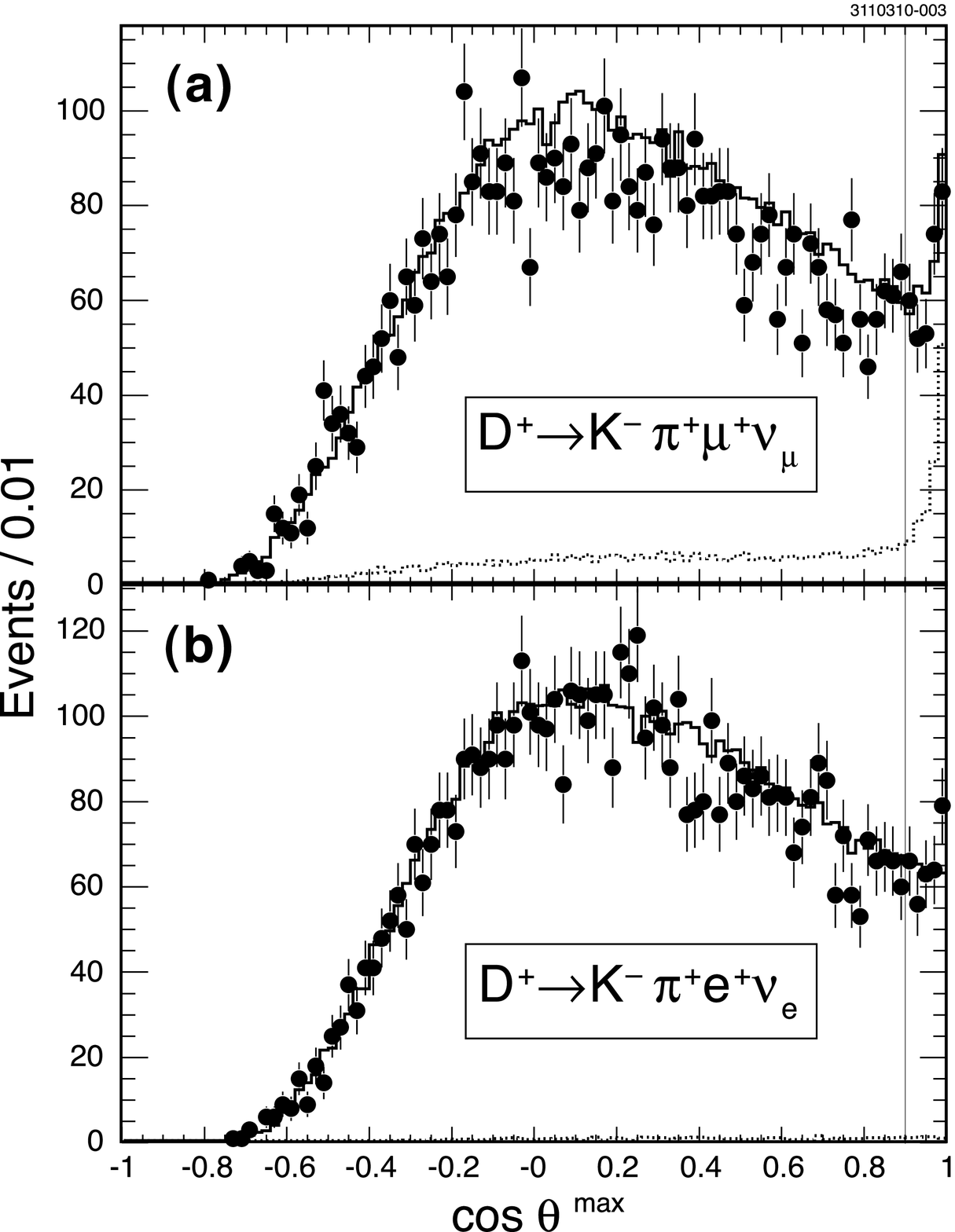}
\caption{Distributions of the largest cosine 
between missing momentum vector and any of the three charged tracks from 
the semileptonic candidate when all cuts are applied except the cut on 
largest cosine. (a) shows the $\cos{\theta^{\rm max}}$ distribution for 
\kpimndk{} candidates, while (b) shows the $\cos{\theta^{\rm max}}$ 
distribution for \kpiendk{} candidates.
We remove all combinations to the right of the vertical line, which removes 
the major part of remaining $K\pi\pi$ background for the semimuonic sample. 
In each plot, the solid histogram shows the signal plus background distribution predicted 
by our Monte Carlo simulation, while the dashed histogram shows the 
predicted background component.
\label{cone}}
\end{figure}

In order to suppress cross-feed from \kpiendk{} decay to our \kpimndk{} sample, 
we construct the squared invariant mass of the lepton candidate, 
$\widetilde{M}_{\mu}^{2} c^4= \left(2E_\textrm{beam}-E_\textrm{Dtag}-E_{{\nu}_{\mu}}-E_{K}-E_{\pi}\right)^2-(cP_{\ell})^2$,
where 
$E_\textrm{Dtag}$ is the reconstructed energy of the $D^-$ produced
against the $\kpilndk{}$ candidate and $E_{K}$, $ E_{\pi}$, $P_{\ell}$ are 
the reconstructed kaon energy, pion energy, and lepton momentum.  We require 
$0 < \widetilde{M}_{\mu}^2 < 0.020  ~\textrm{GeV}^2/c^4 $ to eliminate
both \kpiendk{} cross-feed  and $D^+{} \rightarrow K^{-}\pi^{+}\pi^{+}\pi^{0}$ decays.
In order to suppress backgrounds to \kpimndk{} from $D^+ \rightarrow K^- \pi^+ \pi^+$ 
decays with an accompanying bremsstrahlung
photon, we require that cosine of the minimum angle between three charged tracks and the 
missing momentum direction be less than 0.90. This requirement is illustrated in Fig.~\ref{cone}.

\begin{figure}[htbp]
\includegraphics[width=2.8in]{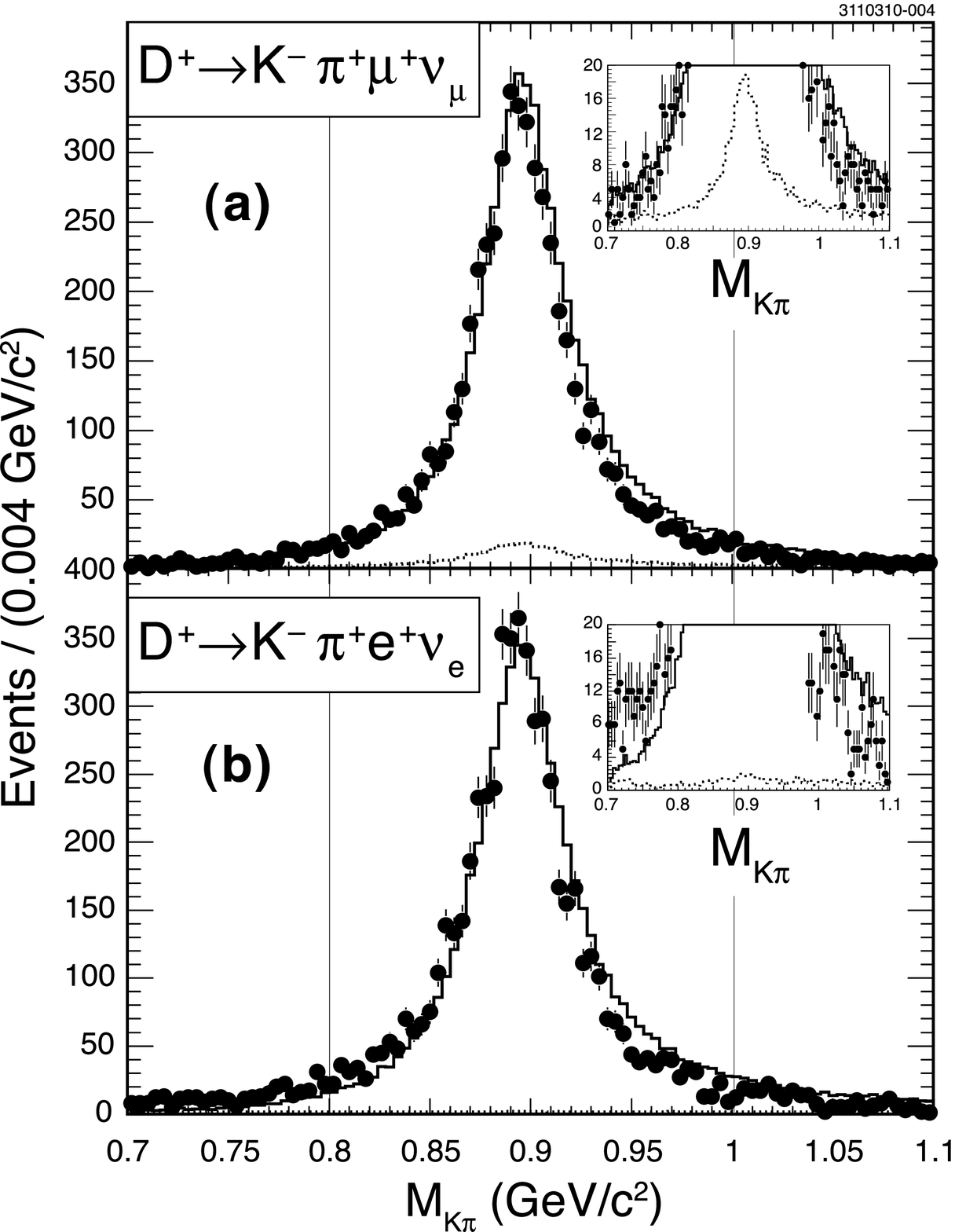}
\caption{ The \mkpi{} distributions for events satisfying our nominal
\kpilndk{} selection requirements. 
(a) shows the \mkpi{} distribution for \kpimndk{} candidates, while 
(b) shows the \mkpi{} distribution for \kpiendk{} candidates.
Over the full displayed mass range, there
are 11\,801 (6\,227 semielectric and 5\,574 semimuonic) events satisfying 
our nominal selection. For this analysis, we use a restricted mass range from 
0.8~--~1.0~\gevcsq{}, which is the region between the vertical lines. 
In each plot, the solid histogram shows the signal plus background distribution predicted 
by our Monte Carlo simulation, while the dashed histogram shows the predicted 
background component. In this restricted region, there are
10865 ( 5\,658 semielectric and 5\,207 semimuonic) events.  The inserted figures are on 
a finer scale to better show the estimated background contributions. 
\label{signal}}
\end{figure}

We obtain 11801 \kpilndk{} candidates.  The \mkpi{} distribution for these \kpilndk{} 
candidates is shown in Fig.~\ref{signal}. Finally, we require
$0.8~\gevcsq{} \le \mkpi \le 1.0~\gevcsq{}$ and select 10865 events.

Two types of Monte Carlo simulations are used throughout this analysis.
The \emph{generic} Monte Carlo simulation is a large charm Monte Carlo sample consisting of 
generic $D\overline{D}$ decays, 
which is primarily used in this analysis to simulate the properties of
backgrounds to our \kpilndk{}
signal states. The \emph{generic} Monte Carlo events are generated by 
\textsc{EvtGen}~\cite{evtgen} and the
detector is simulated using a \textsc{GEANT}-based~\cite{geant} program. 
In much of the form-factor work, we use an SPD Monte Carlo simulation based
on the SPD model described in Sec.~\ref{intro} and summarized by 
Eqs.~(\ref{amp1}--\ref{SPDpole}). We  use
the SPD parameters of Ref.~\cite{formfactor}, \rvee{} =1.504 , \rtwo{} =0.875, 
and we set \rthree{}=0. 

The background shapes in Fig.~\ref{signal} are obtained
using \emph{generic} Monte Carlo simulations.  
Our simulation predicts a 6.5 \% background for our \kpimndk{} sample with 4\% due
to misidentified \kpiendk{} cross-feed events and the rest due to various charm decays. 
The simulation also predicts a 1\% background to our \kpiendk{} sample
with 0.03 \% due to \kpimndk{} cross-feed. 

\mysection{ABSOLUTE AND RELATIVE BRANCHING FRACTIONS}
We have measured both the semimuonic to semielectric relative branching ratio and 
the \krzendk{} and \krzmndk{} absolute branching fractions, which we will denote as ${\mathcal B}_e$ and ${\mathcal B}_\mu$, respectively.
The \brat{} relative branching ratio is expected to be less than 1 due to the reduced phase 
space available to the semimuonic
decay relative to the semielectric decay. The mass-suppressed terms in Eq.~(\ref{amp1}) will change
the relative branching ratio compared to the phase space ratio. 
In the context of the SPD
model, Eq.~(\ref{SPDpole}), the relative branching fraction will depend on $r_3 \equiv A_3(0)/A_1(0)$, which controls the 
strength of the \HT{} form factor and is essentially unknown. It is expected that \brat{}
will increase with increasing values of $r_3$.  

In order to obtain the semimuonic to semielectric branching ratio, we write the observed 
${ D^{+}\rightarrow \bar{K}^{*0}\mu^{+}\nu_{\mu}}$ 
and ${ D^{+}\rightarrow \bar{K}^{*0}e^{+}\nu_{e} }$  yields as

\begin{equation}
%\[
\left(
\begin{array}{c}
y_{e} \\
y_{\mu} \end{array}
\right)
=\left(
\begin{array}{cc}
\epsilon_{e}(\vec{f}\,) & c_{\mu}(\vec{f}\,) \\
       c_{e}(\vec{f}\,) & \epsilon_{\mu}(\vec{f}\,) \end{array}
\right)
%\times\left(
\left(
\begin{array}{c}
n_{e} \\
n_{\mu} \end{array}
\right)
+\left(
\begin{array}{c}
b_{e} \\
b_{\mu} \end{array}
\right),
%\]
\label{eq:BrRatio-yield}
\end{equation}
where $y_{\mu,e}$ are the observed yields, $ b_{\mu,e}$ are 
non-semileptonic backgrounds, and $n_{\mu,e}$ 
give the number of produced semileptonic decays in our data sample.  
The cross-feed matrix, which multiplies the $n_e$ and $n_\mu$  signal vector, 
is constructed from
$ \epsilon_{\mu,e}(\vec{f}\,)$, which are the 
$D^{+} \rightarrow \bar{K}^{*0}\mu^{+}\nu_{\mu}$ and 
$D^{+}\rightarrow \bar{K}^{*0}e^{+}\nu_{e}$ detection
efficiencies, and $ c_{\mu,e}(\vec{f}\,)$, which are the cross-feed efficiencies. For example,
$ c_{\mu}(\vec{f}\,)$  is the efficiency for reconstructing a
$D^{+} \rightarrow \bar{K}^{*0}\mu^{+}\nu_{\mu}$ event as a $D^{+}\rightarrow \bar{K}^{*0}e^{+}\nu_{e}$
candidate.
The $y_{\mu,e}$ yields are obtained
by counting the number of semimuonic and semielectric events in our mass range
$0.8 < \mkpi < 1.0~\gevcsq{}$. The relative branching ratio is given by
$\brat{} = n_\mu/n_e$. 

The vector $\vec{f}$ represents parameters
that the efficiencies and cross-feeds can depend on such as the SPD parameters: 
\rvee{} , \rtwo{}, and \rthree{} 
and the $s$-wave amplitude and phase.
The detection efficiencies, 
$\epsilon_{\mu,e}(\vec{f}\,)$, and the cross-feed efficiencies, $c_{\mu,e}(\vec{f}\,)$, 
were obtained using our Monte Carlo simulations. We will refer to the use of 
Eq.~(\ref{eq:BrRatio-yield}) to obtain
the relative branching ratio, \brat{}, as the \emph{cross-feed method}.

We used the double-tag technique, described in Ref.~\cite{HadronicBrFraction}, 
to measure the 
${D^{+}\rightarrow \bar{K}^{*0}\mu^{+}\nu_{\mu}}$ and 
${D^{+}\rightarrow \bar{K}^{*0}e^{+}\nu_{e} }$  absolute semileptonic 
branching fractions (${\cal B}_{\mu,e}$). We define single tag (ST) events as events where the 
${D^-}$ was fully reconstructed against one of our six tag modes
without any requirement on the recoil ${D^+}$.

\begin{figure}[htbp]
\begin{center}
\includegraphics[width=3.2in]{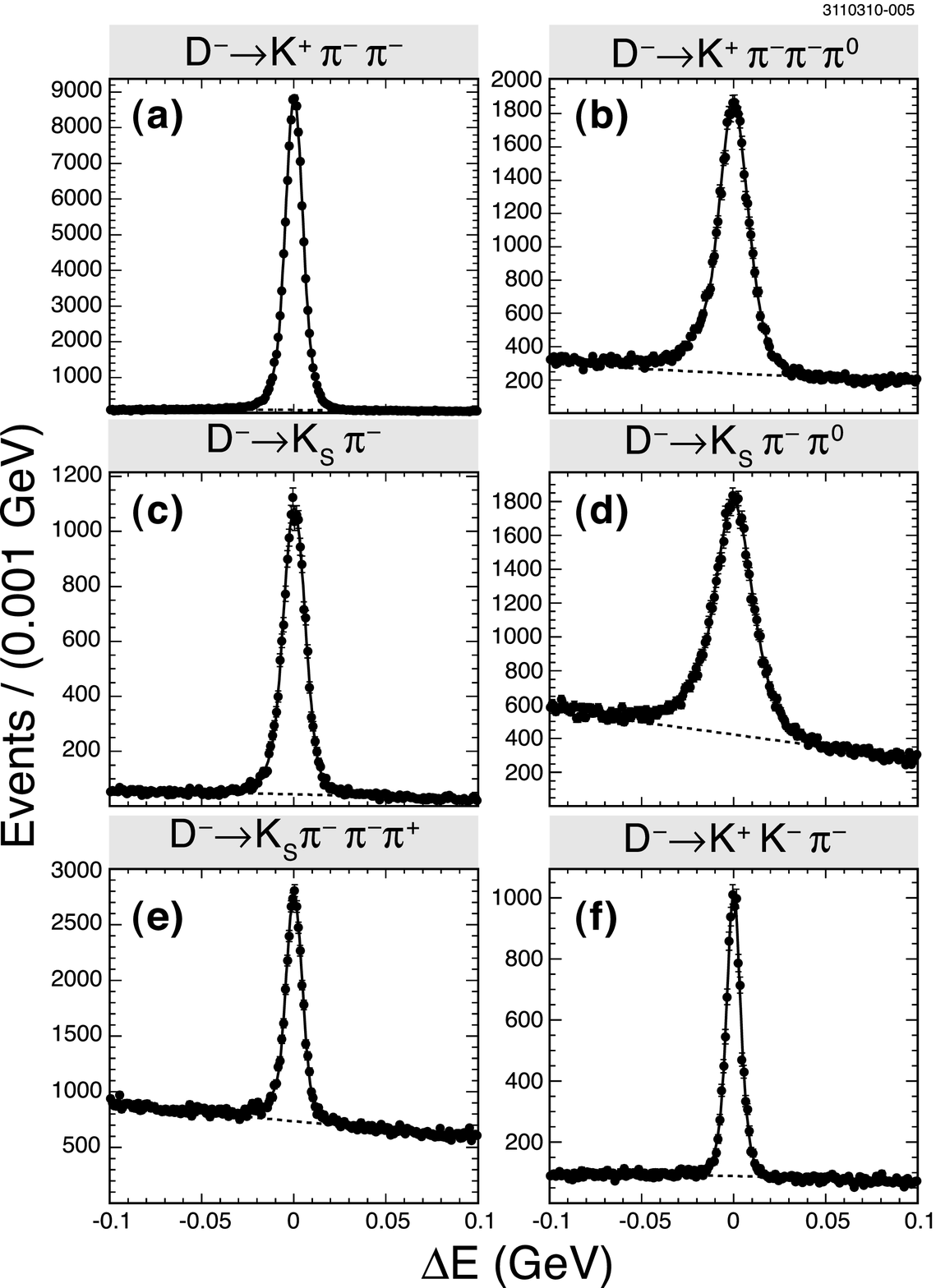}
\caption{Distribution of $\Delta E$ for single tag $D^-$ candidates when $D^+$ and $D^-$ candidates 
have been combined. The distribution for each of the six tags is shown in (a)--(f). 
The points with error bars are the reconstructed yield from the data sample and the curves 
show our fit to the signal peak over the dashed background line.}
\label{fig:abs-st-data}
\end{center}
\end{figure}

We estimate the number of ST events by fitting the $\Delta E$ distributions, shown 
in Fig.~\ref{fig:abs-st-data}, using a binned maximum likelihood fit\footnote{ 
Our fitting function is a sum of Gaussian and Crystal Ball line-shape functions \cite{HadronicBrFraction} 
over a first order Chebyshev background polynomial.}. Here $\Delta E \equiv  E_\textrm{D} - E_\textrm{beam}$, 
where $E_\textrm{D}$ is the energy of the D-tag candidate.

The total number of
reconstructed ${D^-}$  ST events is then
\begin{equation}
\label{eq:ST}
{n_\textrm{ST}^{i} = N_{D^+D^-}\epsilon_\textrm{ST}^{i} {\cal B}_\textrm{tag}^{i}},
\end{equation}
where $ n_\textrm{ST}^{i}$ is the number of ST reconstructed events in the 
$i$-th mode, 
${  N_{D^+D^-}}$ is the number of produced ${ D^+D^-}$  
events in our data sample, ${ \epsilon_\textrm{ST}^{i}}$ is the ST detection 
efficiency, and ${\cal B}_\textrm{tag}^{i}$ is the tag mode branching fraction.

For double tag (DT) events, we reconstruct ${ D^-}$ 
into one of our six tagging modes, and require the presence of 
either a ${ D^{+}\rightarrow \bar{K}^{*0}\mu^{+}\nu_{\mu}}$ or
${ D^{+}\rightarrow \bar{K}^{*0}e^{+}\nu_{e} }$ candidate.
The DT yields are then
\begin{equation}
\label{eq:DT-el}
{n_\textrm{DT}^{e,i}  = N_{D^+D^-}\left(\epsilon_\textrm{DT}^{e,i} {\cal B}_\textrm{tag}^{i}{\cal B}_{e} + 
                      c_\textrm{DT}^{\mu,i} {\cal B}_\textrm{tag}^{i}{\cal B}_{\mu}\right) }
\end{equation}
and
\begin{equation}
\label{eq:DT-mu}
{n_\textrm{DT}^{\mu,i}= N_{D^+D^-}\left(c_\textrm{DT}^{e,i} {\cal B}_\textrm{tag}^{i}{\cal B}_{e} + 
  \epsilon_\textrm{DT}^{\mu,i} {\cal B}_\textrm{tag}^{i}{\cal B}_{\mu}\right) }, 
\end{equation}
respectively.  The yields ${n_\textrm{DT}^{e,i} }$ and ${n_\textrm{DT}^{\mu,i}}$ represent the 
number of reconstructed DT events in semielectric and semimuonic decay
modes after the background subtraction. The efficiencies
${\epsilon_\textrm{DT}^{e,i} }$ and  ${\epsilon_\textrm{DT}^{\mu,i} }$ are the
DT event detection efficiencies for the semielectric and semimuonic decay 
modes. The cross-feed efficiency ${c_\textrm{DT}^{\mu,i}}$ describes how often 
a semimuonic decay is reconstructed as a semielectric candidate, while the cross-feed efficiency   
${c_{DT}^{e,i}}$ describes how often a  semielectric decay is 
reconstructed as semimuonic candidate. The variables 
${\cal B}_{e}$, ${\cal B}_{\mu}$ are the respective ${D^{+}\rightarrow \bar{K}^{*0}e^{+}\nu_{e} }$
and ${D^{+}\rightarrow \bar{K}^{*0}\mu^{+}\nu_{\mu}}$ branching fractions, which we wish to measure.

Dividing Eq.~(\ref{eq:DT-el}) and Eq.~(\ref{eq:DT-mu}) by Eq.~(\ref{eq:ST}),
we have:
\begin{equation}
%\[
\left(
\begin{array}{c}
{n_\textrm{DT}^{e,i}/ n_\textrm{ST}^{i}}   \\
{n_\textrm{DT}^{\mu,i}/ n_\textrm{ST}^{i}} \end{array}
\right)
=\left(
\begin{array}{cc}
 \epsilon_\textrm{DT}^{e,i}/\epsilon_\textrm{ST}^{i} & c_\textrm{DT}^{\mu,i}/ \epsilon_\textrm{ST}^{i} \\
 c_\textrm{DT}^{e,i}/\epsilon_\textrm{ST}^{i}        & \epsilon_\textrm{DT}^{\mu,i}/\epsilon_\textrm{ST}^{i} \end{array}
\right)
%\times\left(
\left(
\begin{array}{c}
{\cal B}_{e} \\
{\cal B}_{\mu} \end{array}
\right) .
%\]
\label{eq:BrRatio-yield-abs}
\end{equation}

Equation~(\ref{eq:BrRatio-yield-abs}) shows how the branching fractions of 
$D^+ \rightarrow \bar{K}^{*0} e^{+} \nu_{e}$ and 
$D^+ \rightarrow \bar{K}^{*0} \mu^{+} \nu_{\mu}$
semileptonic modes depend on the ratio of the DT and the ST yields, the 
detection 
efficiencies, and the cross-feed efficiencies. 

\begin{figure}[htbp]
\begin{center}
\includegraphics[width=3.2in]{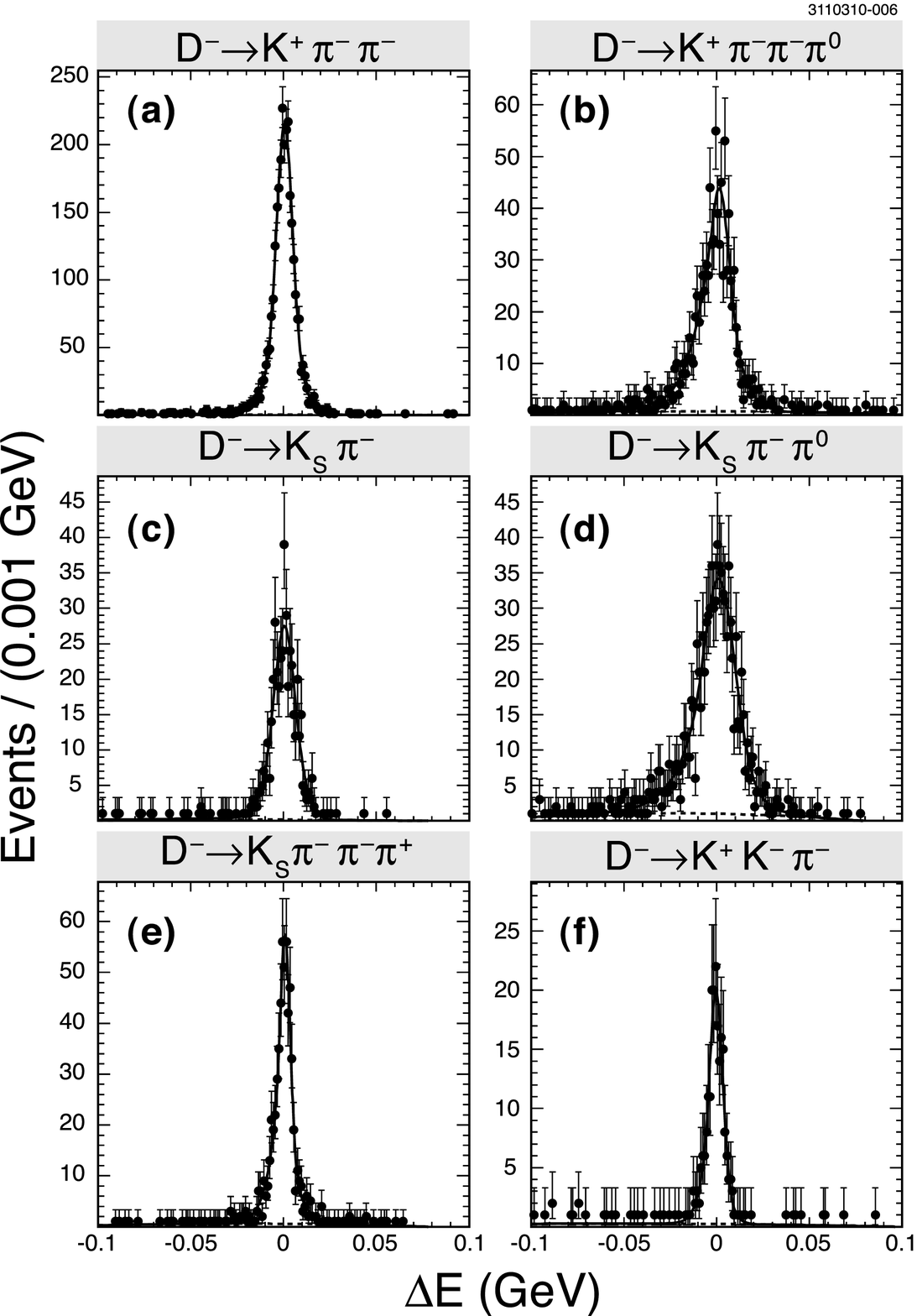}
\caption{Distribution of $\Delta E$ for double tag events for the data, where $D^-$ candidate is
reconstructed in one of the six tag modes [(a)--(f)], and $D^+$ candidate is reconstructed in 
$\bar{K}^{*0} e^{+} \nu_{e}$ mode. The points with error bars are the reconstructed yield from the data 
sample and the curves show our fit to the signal peak over the dashed background line.}
\label{fig:abs-dt-data-el}
\end{center}
\end{figure}

\begin{figure}[htbp]
\begin{center}
\includegraphics[width=3.2in]{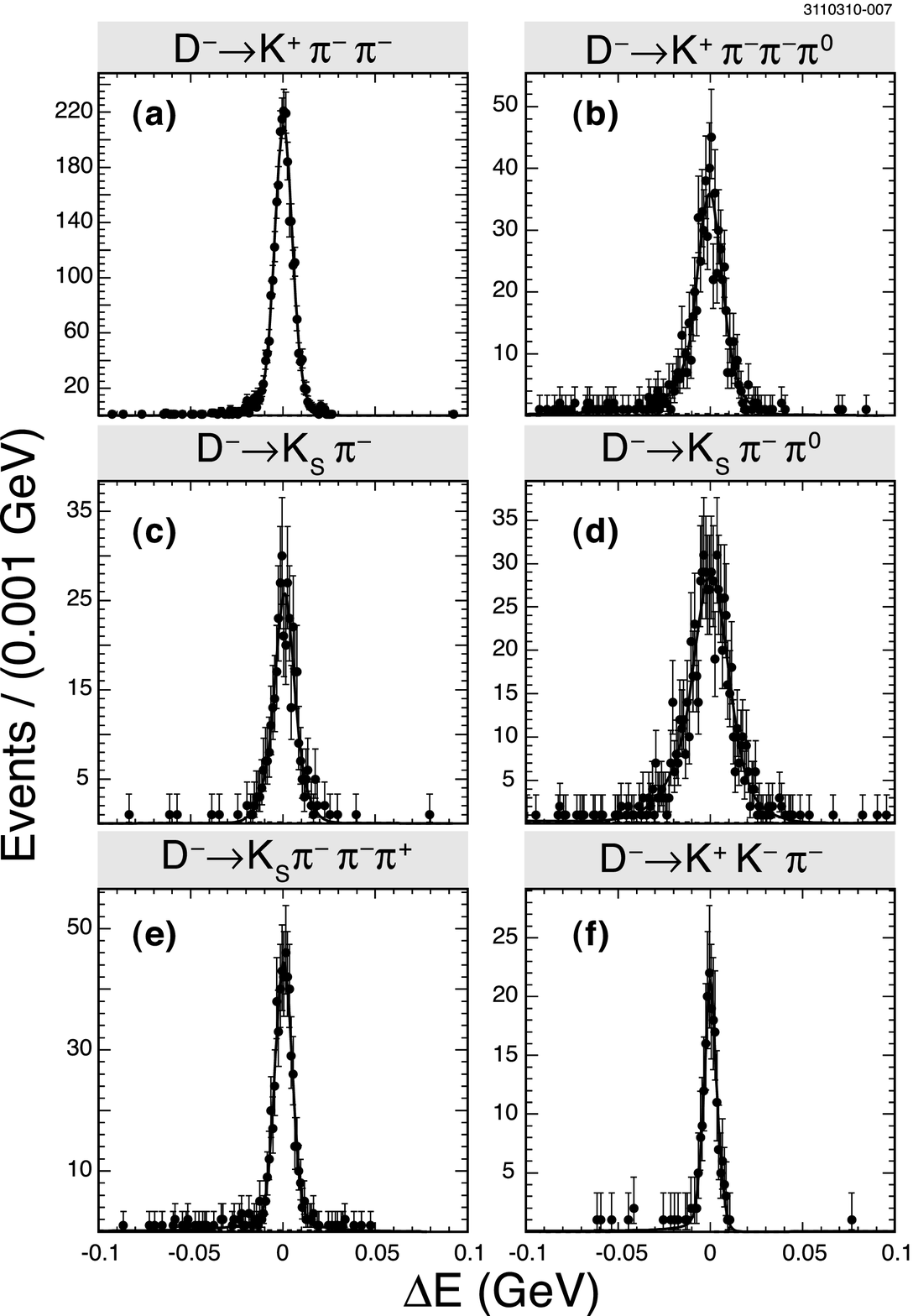}
\caption{Distribution of $\Delta E$ for double tag events, 
where $D^-$ candidate is reconstructed in one of the six tag modes [(a)--(f)], 
and $D^+$ candidate is reconstructed in 
$\bar{K}^{*0} \mu^{+} \nu_{\mu}$ mode. 
The points with error bars are the reconstructed yield from the data sample and the curves 
show our fit to the signal peak over the dashed background line.}
\label{fig:abs-dt-data-mu}
\end{center}
\end{figure}

Figures~\ref{fig:abs-dt-data-el} and~\ref{fig:abs-dt-data-mu} shows 
the $\Delta E$ distributions for our double tag sample. 
For both semileptonic decay modes, about half of our sample comes from the 
$ D^- \rightarrow K^+\pi^-\pi^-$ D-tag mode. The ST yields for this mode 
are nearly background free. 
The cross-feed fraction for the $D^- \rightarrow K^{*0} e^{-} \bar{\nu}_{e}$ 
semileptonic mode is less than 0.02\%, while, for the 
$D^- \rightarrow K^{*0} \mu^{-} \bar{\nu}_{\mu}$ semileptonic
mode, the cross-feed fraction is 3.7\%. The background 
level is about 2.5 times smaller 
for the $D^- \rightarrow K^{*0} e^{-} \bar{\nu}_{e}$ mode than for the
$D^- \rightarrow K^{*0} \mu^{-} \bar{\nu}_{\mu}$ 
mode. The semielectric mode is nearly background free because of 
the effectiveness of the electromagnetic calorimeter, while our semimuonic 
mode uses a variety of less effective kinematic cuts to suppress background 
and cross-feed.

Our absolute branching fraction results are summarized by 
Tables~\ref{tab-abs-br-cond}~and~\ref{tab-abs-br-comparison}. 
Table~\ref{tab-abs-br-cond} gives a ``conditional'' absolute branching
fraction based only on \kpilndk{} decays into
the mass range $0.8 < \mkpi < 1.0 ~\gevcsq{}$. This mass range is required 
for events entering into 
Figs.~\ref{fig:abs-dt-data-el}~and~\ref{fig:abs-dt-data-mu}.  
We find that the total systematic error for the
semielectric and semimuonic absolute branching fractions, presented in
Table I, are comparable.  The dominant systematic error for the
semielectric decay is due to the 1\% uncertainty in the efficiency our
electron identification requirements, while the dominant systematic error
for the semimuonic branching fraction is due to the 0.8\% uncertainty in
the background subtraction. The remaining systematic error, which is 1.2\%
for both the semielectric and semimuonic branching fractions, includes
uncertainties in the final state radiation corrections, as well as
uncertainties in the tracking and particle identification efficiencies for
the kaon and pion tracks. 
Table~\ref{tab-abs-br-comparison}, on the other hand,
relies on models for the \krzb{} line-shape to extrapolate outside 
of the 200 $\rm{MeV}/c^2$ wide mass region where 
our measurements are made in order to report the conventional \krzlndk{} 
absolute branching ratios, which includes events over the entire \mkpi{}
spectrum.  We include an additional, Clebsch-Gordan factor of 1.5
in order to correct for the undetected $\krzb \rightarrow \overline{K}^{0} \pi^0$ decay mode\footnote
{The central values reported in Table~\ref{tab-abs-br-comparison} 
assume that all of the signal events in the $0.8<\mkpi< 1.0~\gevcsq$
mass region, where our $\Delta E$ measurements made, are due to \krzlndk{} decay. 
}.
Finally, we have included an additional $\pm 0.10 ~\%$ contribution 
to the quoted systematic error in Table~\ref{tab-abs-br-comparison}
based on the difference between the \krzb{} extrapolations made using our Generic 
and SPD models. This $\pm 0.10 ~\%$ systematic error contribution includes both distortions to the \krzb{} line shape as well as uncertainties in 
level of non-resonant contributions due to the $s$-wave amplitude.   
\begin{table}[h]
\caption{Conditional absolute branching fractions.
These branching fractions only represent the $K^-\pi^+$ spectrum from $0.8<\mkpi<
1.0~\gevcsq$.}
\label{tab-abs-br-cond}
\begin{center}
\begin{tabular}{cc}
\hline\hline
Mode & Branching fraction [\%] \\
\hline
\condBRe{} & $3.19 \pm 0.04 \pm 0.05 $   \\
\condBRm & $3.05 \pm 0.04 \pm 0.05 $   \\
\hline\hline
\end{tabular}
\end{center}
\end{table}
 \begin{table}[h]
\caption {Comparison of our absolute branching fraction measurements
to previously published data. These branching fractions represent the $K
\pi$ contribution over the full \mkpi{} spectrum and include a systematic
error contribution for uncertainties in the \krzb line shape.}
\label{tab-abs-br-comparison} 
\begin{center}
\begin{tabular}{lcl}
\hline\hline
  & Lumin. [$\rm pb^{-1}$] & ${\mathcal B}_e$ [\%]\\
\hline
These results      & 818 & $5.52 \pm 0.07 \pm 0.13 $         \\
CLEO \cite{absBeCLEO}      & 56 & $5.56 \pm 0.27 \pm 0.23$ \\
World Average \cite{pdg}               & --   & $5.49 \pm 0.31 $         \\
\hline
& Lumin. [$\rm pb^{-1}$] & ${\mathcal B}_\mu$ [\%]\\
\hline
These results  & 818 & $5.27 \pm 0.07 \pm 0.14 $         \\
World Average \cite{pdg}           &--& $5.40 \pm 0.40 $         \\
\hline\hline
\end{tabular}
\end{center}
\end{table}
\begin{table}[htbp]
\caption
{
The  \brat{} branching ratio for the data
based on {\it relative} and {\it absolute} measurements. 
} 
\label{tab-rel-br-fraction-comp}
\begin{center}
\begin{tabular}{ll}
\hline\hline
Method & \brat{}~[\%]\\
\hline
Absolute       & $ 95.98 \pm 1.93 \pm 1.30$  \\
Cross-feed~~~~  & $ 94.64 \pm 1.95 \pm 1.03$ \\
PDG~2008       & $ 98.36 \pm 9.16 $          \\
\hline\hline
\end{tabular}
\end{center}
\end{table}

\begin{figure}[htbp]
\begin{center}
\includegraphics[height=2.5in]{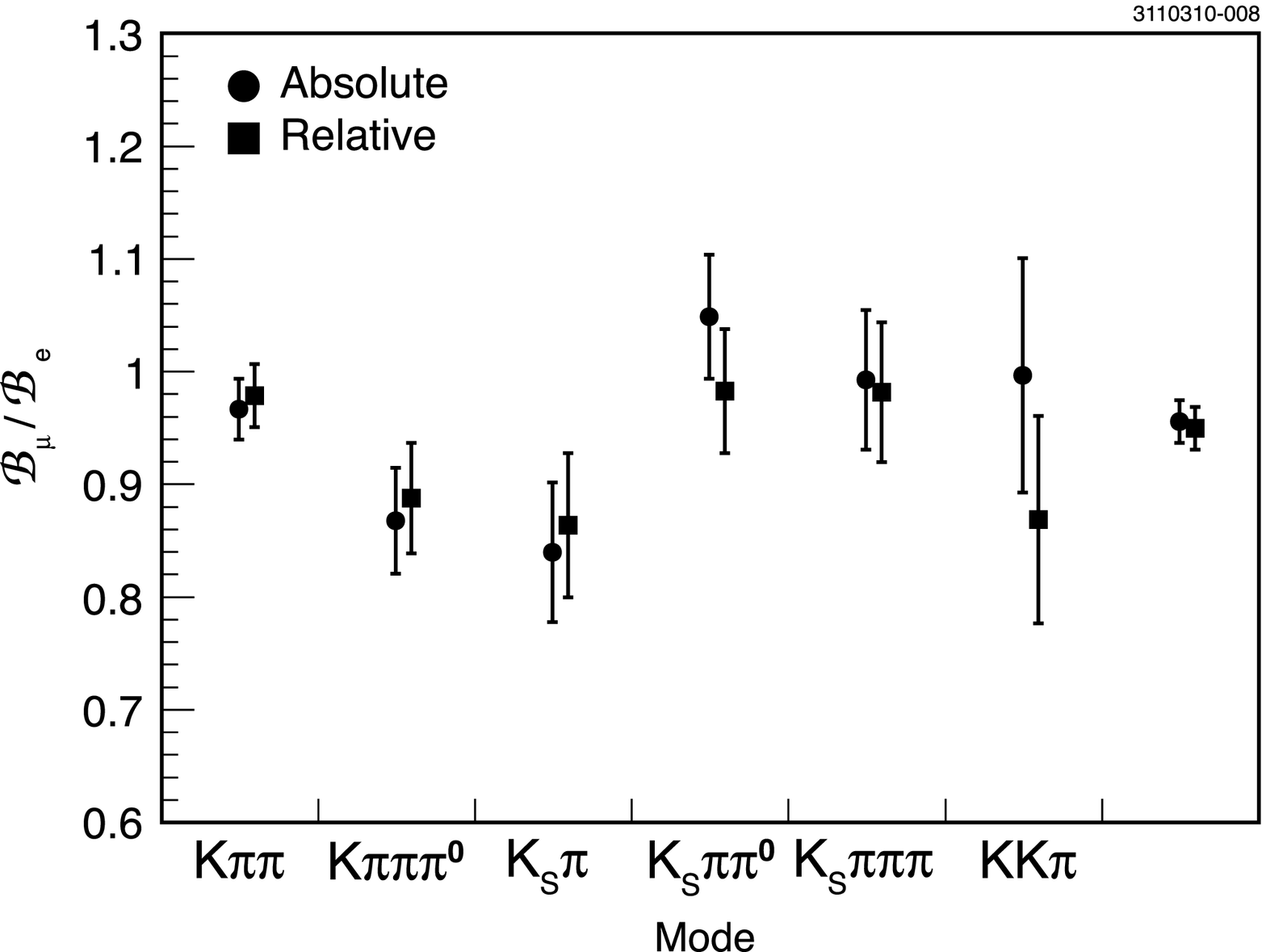}
\caption{Results on the relative branching ratio, \brat{} obtained for the 
six tag states and the error weighted average of these six values. We compare 
the relative branching ratio using the cross-feed method 
[Eq.~(\ref{eq:BrRatio-yield})]  to the ratio of absolute branching fractions.  
Table~\ref{tab-rel-br-fraction-comp} gives a summary of these results.}
\label{fig:relBR}
\end{center}
\end{figure}

Figure~\ref{fig:relBR} and Table~\ref{tab-rel-br-fraction-comp} compare our 
relative \brat{} obtained using the 
cross-feed method to the ratio of absolute branching ratios for the six tag 
states and \emph{generic} and SPD Monte Carlo simulations. The
cross-feed method is reasonably consistent with the ratio of absolute branching fractions.

\mysection{\label{proj}PROJECTIVE WEIGHTING TECHNIQUE} 
We extract the helicity basis form factors using the projective weighting
technique more fully described in Ref.~\cite{helicity-focus}.  For a given
\qsq{} bin, a weight designed to project out a given helicity form
factor, is assigned to the event depending on its \thv{} and \thl{}
decay angles.  We use 25 joint  $\Delta \costhv \times \Delta \costhl$ angular
bins:  5 evenly spaced bins in \costhv{} times 5 bins in \costhl{}.\footnote{When we 
use a combined semielectric and semimuonic sample, we use a 50 component $\vec N$ vector 
with the first 25 angular components reserved for \kpiendk{} candidates and the second 
25 angular components reserved for \kpimndk{} candidates.}  

For each  $q_i^2$ bin, we can write the bin populations $\vec{N}_{i}$ as a sum 
of the expected bin populations $\vec{m}_{\alpha}$ from each,  individual 
form-factor product contribution to Eq.~(\ref{amp1}). Thus $\vec{N}_{i}$ can be written 
as a linear combination  with coefficients $f_{\alpha}(q_{i}^{2})$, 
\begin{eqnarray}
  \vec N_i = f_+(q_i^2)\,\vec m_+ + f_-(q_i^2)\,\vec m_-
            + f_0(q_i^2)\,\vec m_0  \nonumber \\
+ f_I(q_i^2)\,\vec m_I + f_T(q_i^2)\,\vec m_T +f_{TI}(q_i^2)\,\vec m_{TI} \, .
\label{series}
\end{eqnarray}
Each of the six $f_\alpha(q_i^2)$ coefficients is associated with one of 
the form factor products that we wish to measure.
The six $\vec m_\alpha$ vectors are computed 
using SPD Monte Carlo simulations generated with the Eq.~(\ref{amp1}) 
intensity but including just one of the six form factor products. 
For example, $\vec m_+$ is computed using a simulation
generated with an arbitrary function for $H_+ (q^2)$ (such as $H_+ (q^2) = 1$) 
and zero for the remaining five form factors. 
The $f_\alpha(q_i^2)$ functions are proportional to the true
$H^2_\alpha(q_i^2)$ along with multiplicative factors such as
$G_F^2 \left| V_{cs} \right|^2 (q^2 - m_\ell ^2)$
and acceptance corrections.

Reference~\cite{helicity-focus} shows how Eq.~(\ref{series}) can be solved
for the six form factor products $H_+^2(\qsq)$, $H_-^2(\qsq)$, $H_0^2(\qsq)$,
\hoHO{}, \HsT{}, and \HTHO{} by making six weighted $q^2$ histograms. 
The weights are directly constructed from the six $\vec m_\alpha$ vectors.

\begin{figure}[htbp]
\includegraphics[width=3.25in]{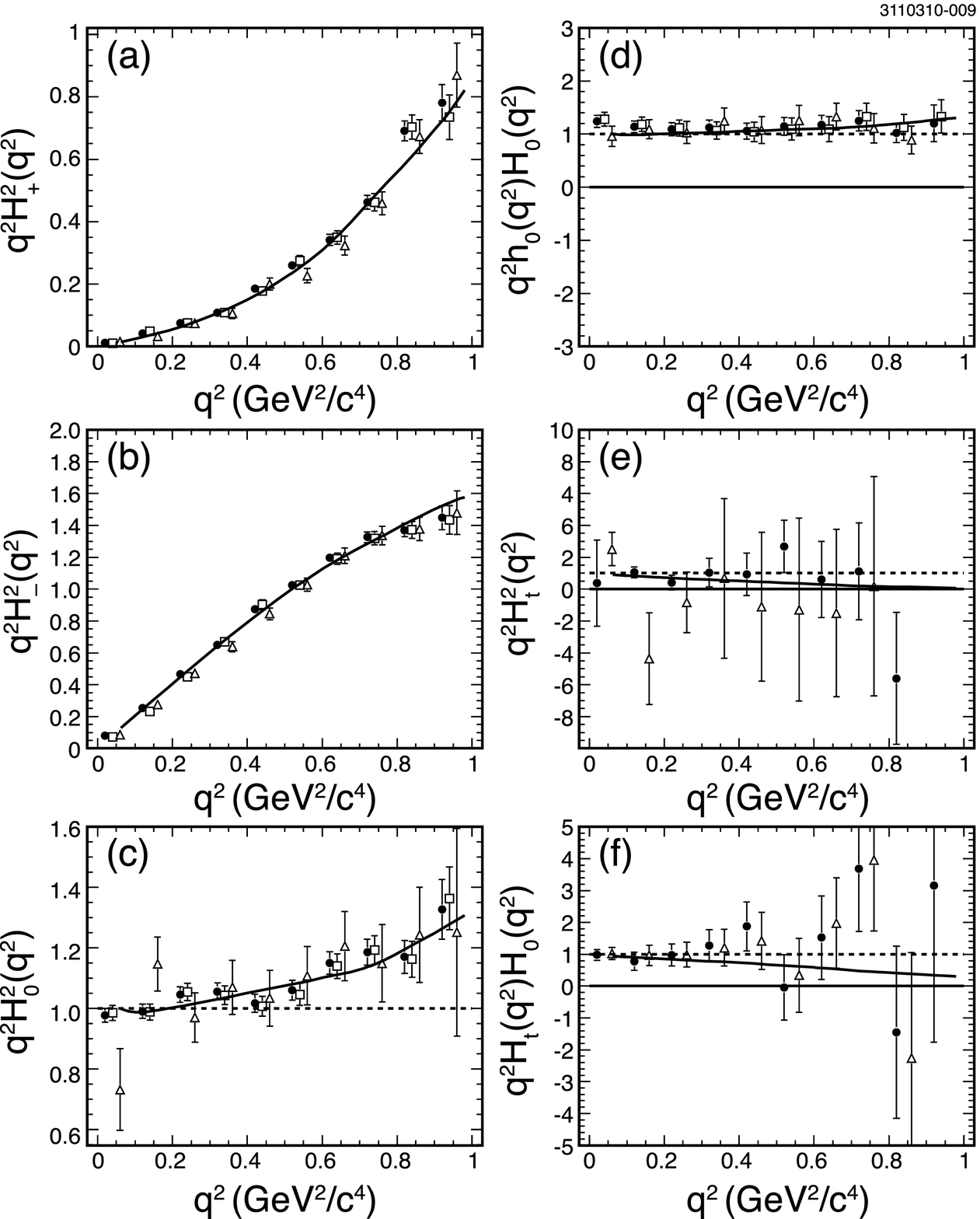}
\caption{
Non-parametric form factor products obtained for the SPD
Monte Carlo sample
(multiplied by \qsq{}) for ten, evenly spaced \qsq{} bins.
The reconstructed form factor products are shown as the points with error bars,
where the error bars represent the statistical uncertainties.
The three points at each \qsq{} value are: 
{\it filled circles} a combined \kpimndk{} \& \kpiendk{} sample, 
{\it empty squares}  \kpiendk{} only,  and  {\it empty triangles} 
\kpimndk{} only. The solid curves represent our SPD model, which was used to 
generate the Monte Carlo sample.
The histogram plots are:
(a)~$\qsq H_+^2(\qsq)$,
(b)~$\qsq H_-^2(\qsq)$,
(c)~$\qsq H_0^2(\qsq)$,
(d)~$\qsq h_0(\qsq)H_0(\qsq)$,
(e)~$\qsq \HsT$, and
(f)~$\qsq \HTHO$.
\label{pop}}
\end{figure} 
Figure~\ref{pop} shows the six form factor products 
multiplied by \qsq{} obtained from a Monte Carlo simulation using our 
selection requirements. Because the isolated \kpiendk{} sample provides no 
useful information on the mass-suppressed form factor products 
\HsT{} and \HTHO{}, the second point is not plotted for these
two form factor products.
The Monte Carlo sample was generated with our SPD Monte Carlo with 
$r_3 = 0$ and was run with three times our data sample. The reconstructed 
form factor products in the 
Monte Carlo simulation are a good match to the input model indicating
that the projective weighting method is reasonably unbiased. 

We turn next to a discussion of our normalization convention.
Equation~(\ref{helicity}) tells us that as $\qsq{} \rightarrow 0$,
$\qsq{} H_\pm^2(\qsq) \rightarrow 0$; and
$\qsq{} \Hszer{}$, $\qsq{} \hoHO{}$, $\qsq{} \HsT$,~$\qsq{} \HTHO{}$
all approch the \emph{same} constant.
Therefore, we normalized the form factor products in Fig.~\ref{pop}
by scaling the weighted histograms
by a \emph{single} common factor so that  $\qsq{} \Hszer{}= 1$ as $\qsq{} \rightarrow 0$ 
based on the $\qsq{} \Hszer{}$ measured in the combined \kpiendk{} and \kpimndk{} sample.  

Figure~\ref{pop} shows that the isolated \kpimndk{} and \kpiendk{} samples produce 
similar error bars for the measured \Hspls{}, \Hsmin{}, and \hoHO{} form factor products, 
while the \Hszer{} errors are much larger for the \kpimndk{} sample
than for the \kpiendk{}.  This is due to the large correlation between the \Hszer{} and  
\HsT{} form factors present in the \kpimndk{} sample owing to the similarity in their 
associated angular distributions. For this reason, the error bars on the \HsT{} form 
factor product are dramatically reduced when one combines the \kpimndk{} and 
\kpiendk{} samples. 

\begin{figure}[htbp]
\begin{center}
\includegraphics[height=1.8in]{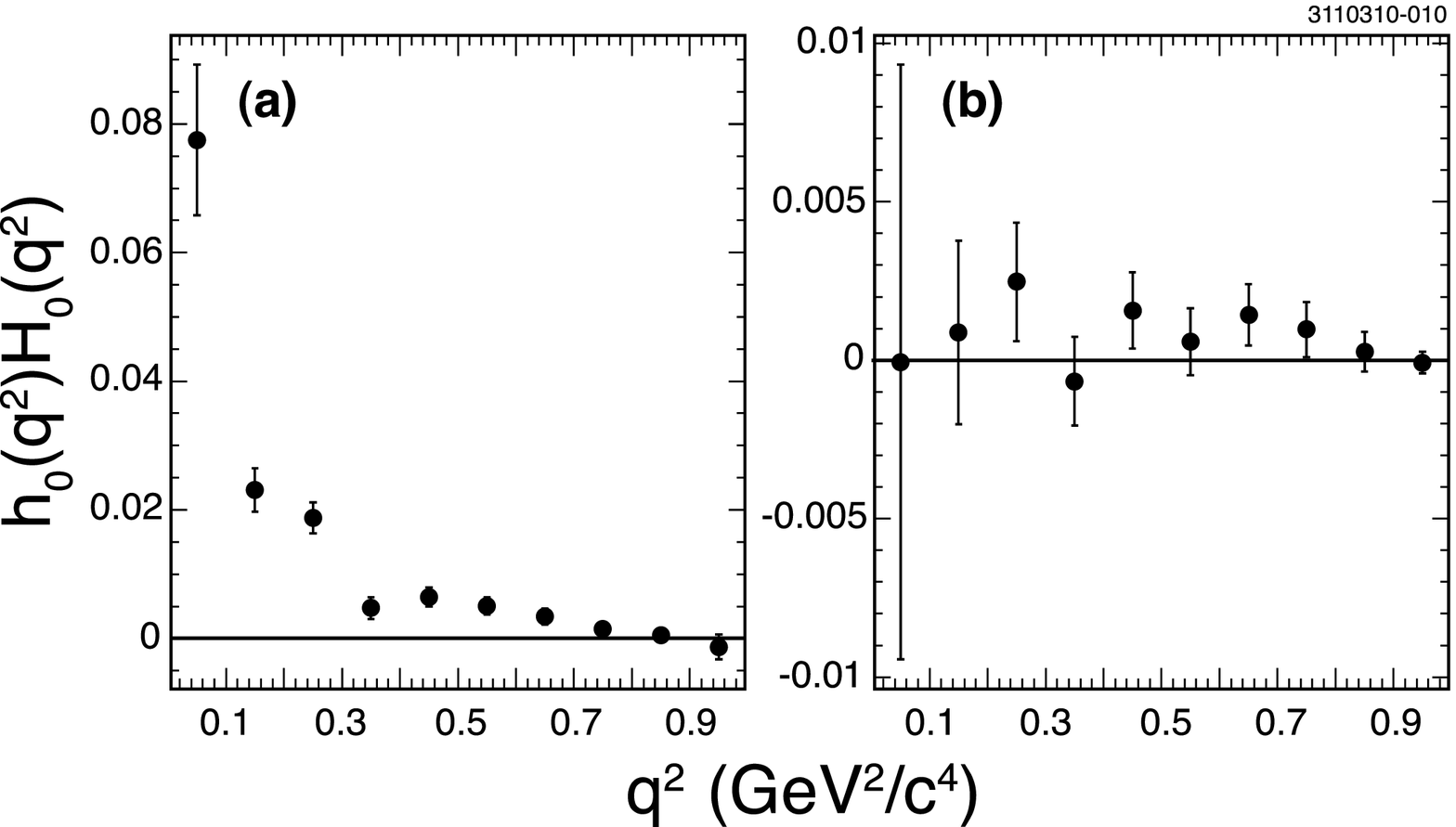}
\end{center}
\caption{
We show uncorrected plots of the $h_0(\qsq)H_0(\qsq)$ for data with \kpimndk{} 
and \kpiendk{} combined. 
(a) is for events below the nominal \krz{} pole:  
$0.8< \mkpi{} < 0.9 ~\gevcsq{}$.  (b) is for 
events above the nominal pole: $0.9< \mkpi{} < 1.0 ~\gevcsq{}$. There is a 
strong $h_0(\qsq)H_0(\qsq)$ 
signal below the nominal pole but no evidence for a non-zero 
$h_0(\qsq)H_0(\qsq)$ form factor above the 
pole. Note the order of magnitude difference in the y-axis scales between 
the left and right plots. 
\label{phase}}
\end{figure}

\begin{figure}[htbp]
\includegraphics[width=3.25in]{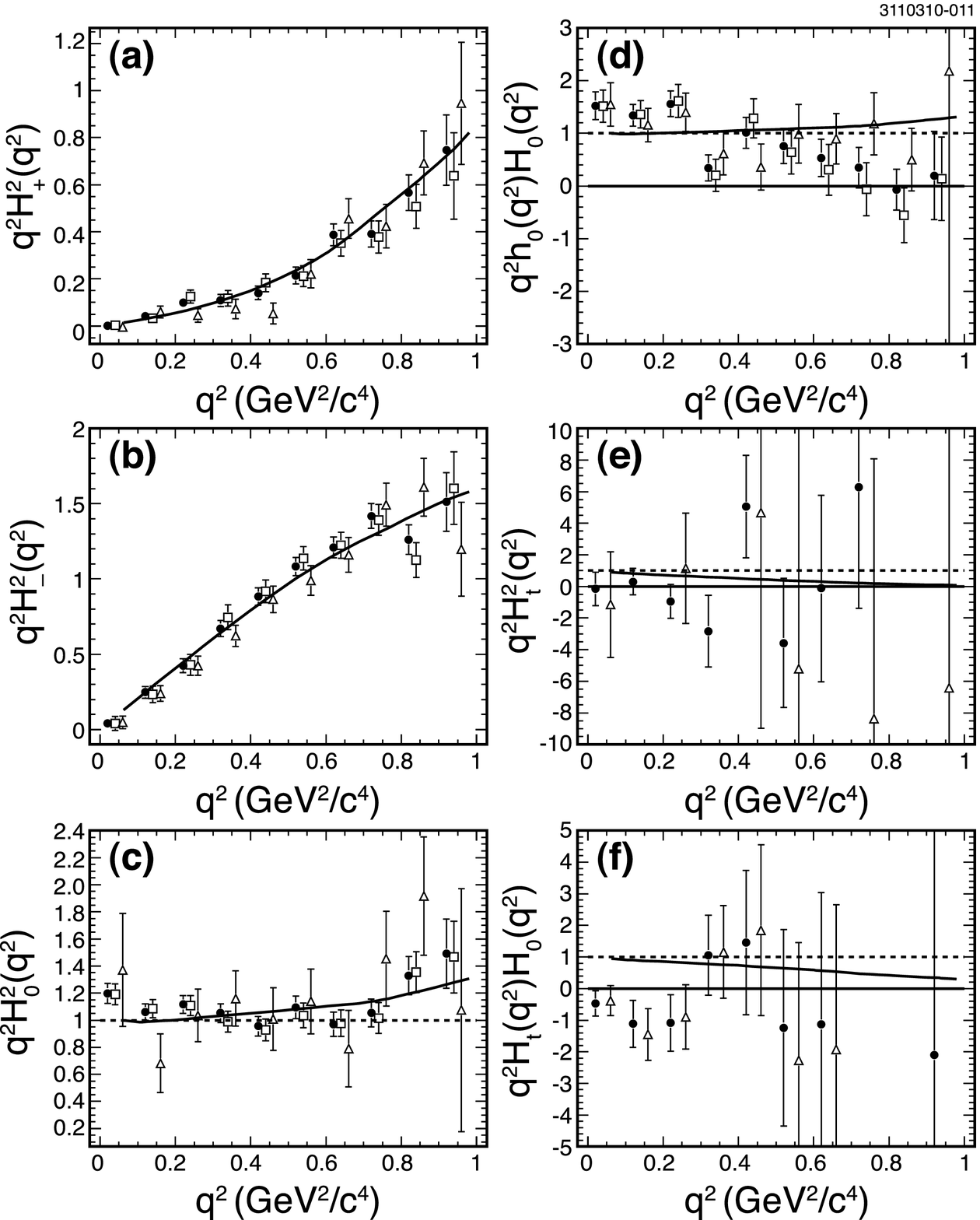}
\caption{
Non-parametric form factor products obtained for the data
(multiplied by \qsq{}) for ten evenly spaced \qsq{} bins.
The reconstructed form factor products are shown as the points with error bars,
where the error bars represent the statistical uncertainties.
The three points at each \qsq{} value are: 
{\it filled circles} a combined \kpimndk{} \& \kpiendk{} sample, 
{\it empty squares}  \kpiendk{} only,  and  {\it empty triangles} 
\kpimndk{} only. 
The solid curves show our SPD model.
The histogram plots are:
(a)~$\qsq H_+^2(\qsq)$,
(b)~$\qsq H_-^2(\qsq)$,
(c)~$\qsq H_0^2(\qsq)$,
(d)~$\qsq h_0(\qsq)H_0(\qsq)$,
(e)~$\qsq \HsT$, and
(f)~$\qsq \HTHO$.
\label{sixnormZ}}
\end{figure}

\mysection{\label{FFresults} FORM-FACTOR RESULTS}
We turn next to a discussion of our form-factor measurements.
Figure~\ref{phase} compares the $h_0(\qsq)H_0(\qsq)$ distribution below the nominal pole 
(a) to that above the nominal \krz{} pole (b). Figure~\ref{phase} shows that
there is no significant $h_0(\qsq)H_0(\qsq)$ signal above the \krz{} pole. The absence of a $h_0(\qsq)H_0(\qsq)$ 
signal above the nominal \krz{} shows that our data are consistent with the $\delta_s$ phase
obtained in Refs.~\cite{helicity-focus,anomaly,formfactor}.
A related interference pattern was observed in the FOCUS \cite{anomaly} discovery  of the $s$-wave interference 
in \kpimundk{} decay. We can thus improve our statistical errors by restricting our $h_0(\qsq)H_0(\qsq)$ 
measurements to events with $0.8< \mkpi{} < 0.9 ~\gevcsq{}$.  This additional requirement was applied to 
the $\qsq h_0(\qsq)H_0(\qsq)$ plot of Fig.~\ref{sixnormZ}, while the other five form factor products use the 
full  $0.8< \mkpi{} < 1.0 ~\gevcsq{}$ mass range. 

Figure~\ref{sixnormZ} shows the six form factor products 
multiplied by \qsq{} obtained for data 
using our $\qsq{} \Hszer{} = 1$ as $\qsq{} \rightarrow 0$ normalization 
convention. The background was subtracted using our Monte Carlo samples.
Although the data are a reasonably good match to the SPD model for the 
$\qsq H_0^2(\qsq)$ and $\qsq H_\pm^2(\qsq)$ form factors, the model does 
not match the data for $\qsq h_0(\qsq)H_0(\qsq)$, and
the mass-suppressed form factors $\qsq \HsT$ and $\qsq \HTHO$. 
These disagreements will be discussed in Sec.~\ref{sum}.

Because of our excellent \qsq{} resolution, there is negligible correlation
among the ten \qsq{} bins for a given form factor product, but the relative
correlations between different form factor products in the same \qsq{}
bin can be much larger. Most of the correlations are
less than 30 \%. There are, however, some very strong ($> 70\%$) correlations for $\Hmin$ 
with various other form factors -- most notably in the three lowest \qsq{} bins in 
the correlations between the \Hmin{} and the $H_T~H_0$ as well as $H_0^2$ form factor
products.

Table~\ref{summary},  a tabular representation of Fig.~\ref{sixnormZ} 
for the \kpimndk{} and \kpiendk{} combined sample,
gives the center of each \qsq{} bin, the measured form-factor product, its statistical 
uncertainty (first error) and
its estimated systematic uncertainty (second error). The biggest source
of the systematic uncertainty is from the background estimation. 
We separately consider systematic uncertainties from non-semileptonic decay backgrounds, 
and semileptonic decay backgrounds. The semileptonic backgrounds include cross-feed as 
well as semimuonic
events where the pion and muon are exchanged. 

For the background uncertainty, we assign a conservative systematic error  
by increasing the level of the non-semileptonic background and semileptonic background subtractions 
by a factor of 1.5 and comparing these form factor products to the results
with the nominal background subtractions. For \Hspls{} and \hoHO{}, the non-semileptonic 
and semimuonic background subtraction systematic uncertainty is
less than 20 \%  of the statistical error, while for the other four form factor products the
systematic error is less than 40\% of the statistical error.

We also assess a relative systematic error due to uncertainties in
track reconstruction and particle identification efficiencies. The systematic
uncertainty from this source is rather small since we are reporting
form factor \emph{shapes} rather than absolute normalization. This uncertainty
is estimated as less than 1.9 \% for all the form factor products.
Finally, we assess a scale error of 13.4\% on the \hoHO{} form factor
product due to the uncertainties in the $A$ and $\delta$ values reported in
Ref.~\cite{formfactor}. When this $s$-wave scale error is added in quadrature
to the subtraction systematic error, the total systematic error rises to about
85\% of the statistical error in the lowest three \qsq{} bins of the \hoHO{}
form factor product, but systematic errors on the form factor shape are less than 20\% of
the statistical error. 

\begin{figure}[htbp]
\begin{center}
\includegraphics[height=2.0in]{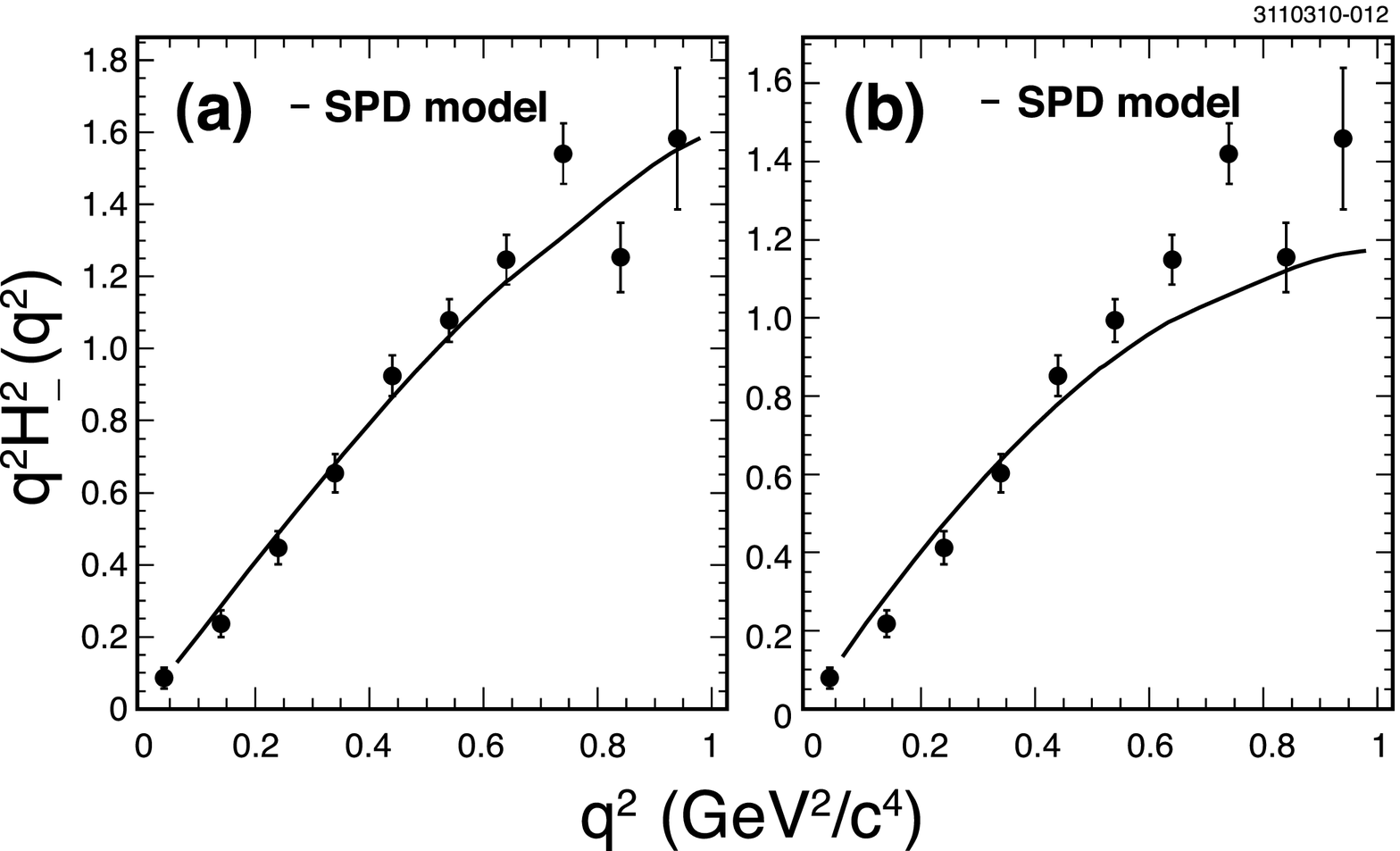}
\end{center}
\caption{
Evidence for finite pole masses.  We show the measured $\qsq \Hsmin{}$ form factor shown 
in Fig.~\ref{sixnormZ} overlayed with two models. 
(a) uses the same SPD model shown in Fig.~\ref{sixnormZ} while (b) overlays the data with a SPD model where
the axial and vector poles [$M_A$ and $M_V$ in Eq.~(\ref{amp1})] are set to infinity. 
We show the data with \kpimndk{} and \kpiendk{} combined. The slight scale difference between the data points in 
the two plots is an 
artifact of our $\qsq{} \Hszer{} = 1$ as $\qsq{} \rightarrow 0$ normalization scheme, which is based on the two 
different pole mass predictions for  the \Hszer{} form factor product. 
\label{finite}} 
\end{figure}

\begin{table*}[htp]
\caption{Summary of form factor product results for ten, evenly spaced \qsq bins for the \kpimndk{} and \kpiendk{} combined sample.
The first and second errors are statistical and systematical uncertainties,
respectively. The numbers are normalized using the condition:
$\qsq{} \Hszer{} = 1$ as $\qsq{} \rightarrow 0$.}
\begin{ruledtabular}
\begin{tabular}{cccccc}
 \qsq{} & \qsq{} \Hspls{}&  \qsq{} \Hsmin{}& \qsq{} \Hszer \\ 
\hline
0.05&0.0013$\pm$0.0061$\pm$0.0010&0.0398$\pm$0.0304$\pm$0.0099&1.1979$\pm$0.0737$\pm$0.0276& \\ 
0.15&0.0417$\pm$0.0135$\pm$0.0026&0.2467$\pm$0.0380$\pm$0.0146&1.0598$\pm$0.0616$\pm$0.0253& \\ 
0.25&0.0993$\pm$0.0209$\pm$0.0036&0.4242$\pm$0.0471$\pm$0.0221&1.1160$\pm$0.0656$\pm$0.0274& \\ 
0.35&0.1079$\pm$0.0259$\pm$0.0039&0.6704$\pm$0.0535$\pm$0.0175&1.0520$\pm$0.0690$\pm$0.0217& \\ 
0.45&0.1401$\pm$0.0290$\pm$0.0031&0.8822$\pm$0.0575$\pm$0.0120&0.9556$\pm$0.0721$\pm$0.0203& \\ 
0.55&0.2140$\pm$0.0358$\pm$0.0026&1.0809$\pm$0.0605$\pm$0.0025&1.0941$\pm$0.0832$\pm$0.0181& \\ 
0.65&0.3874$\pm$0.0457$\pm$0.0057&1.2094$\pm$0.0692$\pm$0.0017&0.9692$\pm$0.0891$\pm$0.0165& \\ 
0.75&0.3907$\pm$0.0548$\pm$0.0060&1.4181$\pm$0.0830$\pm$0.0085&1.0531$\pm$0.1030$\pm$0.0195& \\ 
0.85&0.5670$\pm$0.0759$\pm$0.0090&1.2612$\pm$0.0982$\pm$0.0164&1.3298$\pm$0.1415$\pm$0.0307& \\ 
0.95&0.7475$\pm$0.1495$\pm$0.0084&1.5113$\pm$0.1952$\pm$0.0263&1.4912$\pm$0.2539$\pm$0.0421& \\ 
\hline
\qsq{} & \qsq{} \hoHO& \qsq{} \HsT{}&  \qsq{}\HTHO \\ 
\hline
0.05&1.5263$\pm$0.2649$\pm$0.2068&-0.1535$\pm$1.0530$\pm$0.2330&-0.4717$\pm$0.4033$\pm$0.1983& \\ 
0.15&1.3410$\pm$0.2081$\pm$0.1802&0.3069$\pm$0.8381$\pm$0.3261&-1.1157$\pm$0.7390$\pm$0.3345& \\ 
0.25&1.5601$\pm$0.2470$\pm$0.2092&-0.9425$\pm$1.0708$\pm$0.4993&-1.0842$\pm$0.8925$\pm$0.2879& \\ 
0.35&0.3432$\pm$0.2450$\pm$0.0657&-2.8312$\pm$2.2685$\pm$1.1741&1.0604$\pm$1.2657$\pm$0.3050& \\ 
0.45&1.0085$\pm$0.2927$\pm$0.1378&5.0488$\pm$3.2535$\pm$1.3110&1.4500$\pm$2.2843$\pm$0.5273& \\ 
0.55&0.7593$\pm$0.3344$\pm$0.1186&-3.5770$\pm$4.0787$\pm$1.6076&-1.2391$\pm$3.1060$\pm$0.3136& \\ 
0.65&0.5340$\pm$0.3524$\pm$0.0906&-0.1290$\pm$5.8905$\pm$2.2112&-1.1319$\pm$4.1718$\pm$0.2507& \\ 
0.75&0.3474$\pm$0.3856$\pm$0.0758&6.2982$\pm$7.6928$\pm$2.1522&9.9457$\pm$7.8013$\pm$0.7991& \\ 
0.85&-0.0682$\pm$0.3905$\pm$0.0538&-16.9593$\pm$10.8847$\pm$3.1543&-13.1707$\pm$11.6553$\pm$0.0672& \\ 
0.95&0.1968$\pm$0.8383$\pm$0.0266&-75.1674$\pm$33.6395$\pm$4.8926&-2.1058$\pm$16.0185$\pm$0.0680& \\
\end{tabular}
\end{ruledtabular}
\label{summary}
\end{table*}
Figure~\ref{finite} illustrates our sensitivity to the pole masses in Eq.~(\ref{helicity}) by comparing measurements
of the $\qsq~\Hsmin$ form factor product to a model with spectroscopic axial and vector pole masses versus a 
model with infinite pole masses, implying \emph{constant} axial and vector form factors.  Our data favor the 
spectroscopic pole masses given in Eq.~(\ref{SPDpole}), for the high \qsq{} bins of the \Hsmin{} form 
factor product.  The other five form factor products are consistent with either pole mass choice.  

\begin{figure}[htbp]
\includegraphics[height=2.0in]{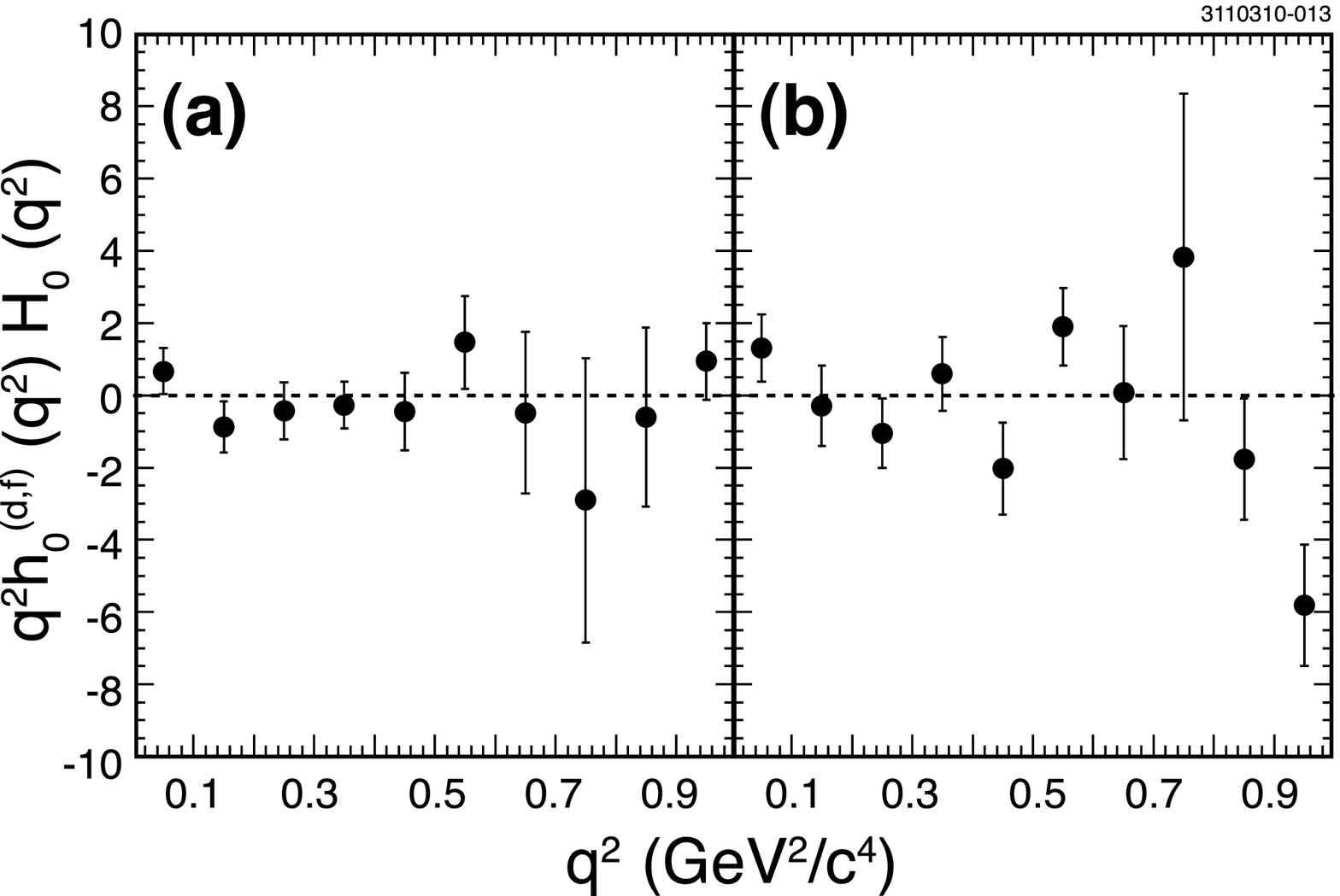}
\caption{
Measurements of the  $d$-wave form factor product (a) and $f$-wave form factor product (b) for an assumed 
phase of 0 radians relative to the \krzb{} Breit-Wigner amplitude. 
\label{BESTFD}}
\end{figure}

It is of interest to search for the possible
existence of additional non-resonant amplitudes of higher angular momentum. 
It is fairly simple to extend Eq.~(\ref{amp1}) to account for potential
$d$-wave or $f$-wave interference with the \krzb Breit-Wigner amplitude.
We search specifically for a possible zero helicity $d$-wave or $f$-wave
piece that interferes with the zero helicity \krzb{} contribution. One expects
that such potential $h^{(d)}_0(q^2)$ and $h^{(f)}_0(q^2)$ form factors would
peak as $1/\sqrt{\qsq}$ near $\qsq{} \rightarrow 0$ as is the case for the
other zero helicity contributions \Hzer{} and \hzer{}. If so, the zero helicity
contributions should be much larger than potential $d$- or $f$-wave $\pm 1$
helicity contributions. 
The $d$-wave projectors are based on an additional interference 
term of the form
\begin{equation}
4\,\sinthlsq (3\,\cos^2 \thv - 1)\,H_0(q^2)\,h^{(d)}_0(q^2)\,
{\mathop{\mathrm{Re}}\nolimits}\{A_de^{-i\delta_d} \bw\}.
\end{equation}
To search for the presence of zero helicity $d$-wave amplitude 
we use the technique of Ref.~\cite{helicity-focus} to construct a 
projector which is orthogonal to the projectors for each 
of the six terms in Eq.~(\ref{amp1}).
The $f$-wave weights are based on an additional interference term of the form
\begin{equation}
4\,\sinthlsq (5\,\cos^3 \thv - 3 \cos \thv)\,H_0(q^2)\,h^{(f)}_0(q^2)\,
{\mathop{\mathrm{Re}}\nolimits}\{A_fe^{-i\delta_f} \bw \}.
\end{equation}
Averaging over the Breit-Wigner intensity, the interference should be proportional to
$A_{d,f}\,\sin{\delta_{d,f}}\,\HDFint{}$ and will disappear 
when the non-resonant amplitude 
is orthogonal to the average, accepted \krzb{} amplitude.  
Fig.~\ref{BESTFD} shows the $\qsq{} h_0^{(d,f)}(\qsq) H_0 (\qsq)$ form factor
products obtained in the data using projective weights generated assuming a phase of zero. The projective
weights are normalized so that $\qsq{} h^{(d,f)}(\qsq) H_0 (\qsq) = 1 $ in the limit $\qsq \rightarrow 0$
if the putative $d$,$f$ -wave amplitude had the same strength  as the $s$-wave amplitude relative to the 
\krzb{} Breit-Wigner amplitude. 
There is no evidence for either a $d$-wave or $f$-wave component with this phase.   

\begin{figure}[htbp]
\includegraphics[height=2.2in]{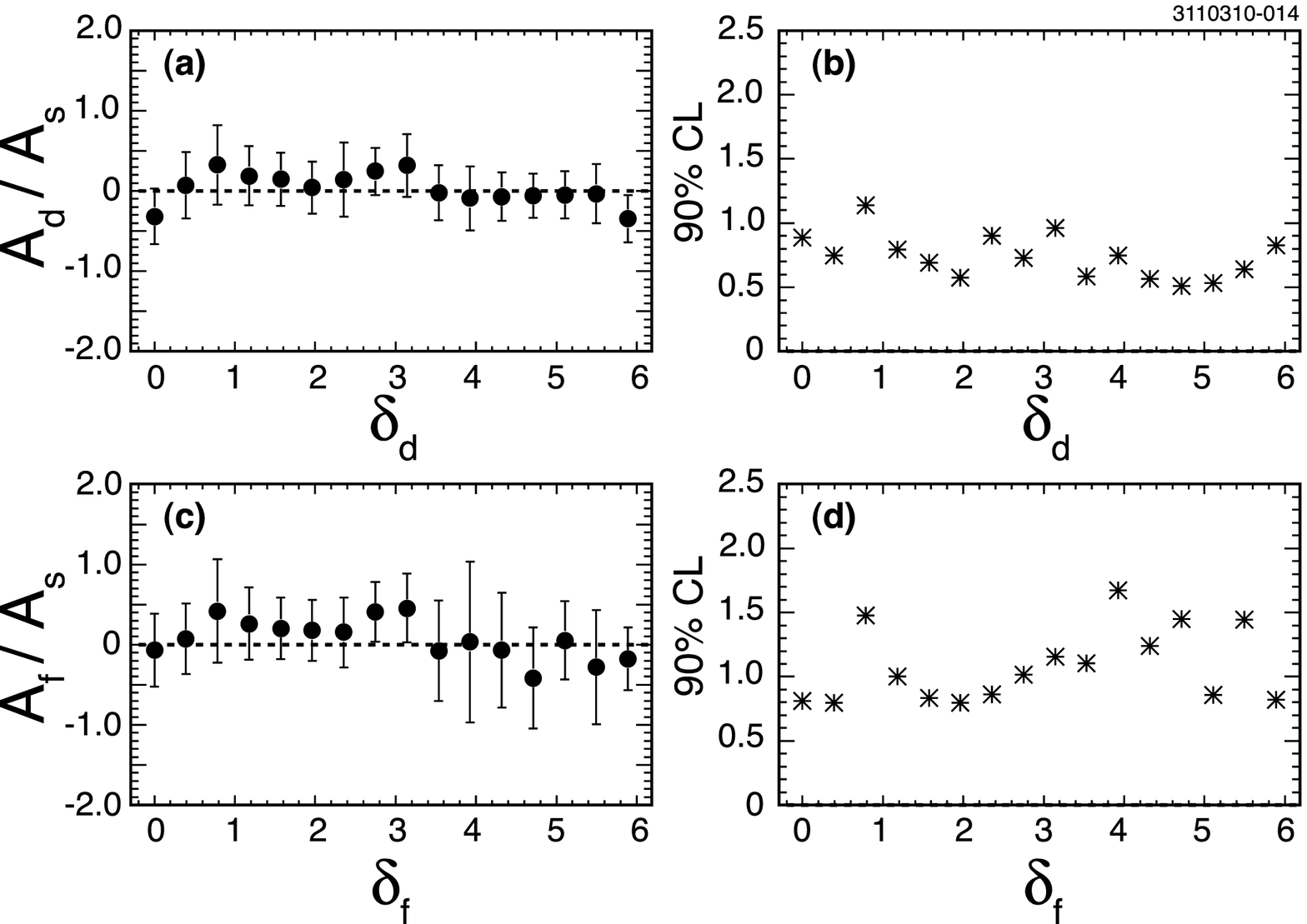}
\caption{Search for d-wave, (a)~and~(b), and f-wave, (c)~and~(d), interference 
effects for each phase assumption as described in the text. The phases 
$\delta_d$ and $\delta_f$ represent the phase of possible d and f -wave 
contributions relative to the phase of the \krzb Breit-Wigner amplitude. 
They are measured in radians.}
\label{wavelimits}
\end{figure}
Figure~\ref{wavelimits} shows our amplitude and limits for sixteen phase assumptions.  
As illustrated by Fig.~\ref{phase}, our ability to measure a non-resonant 
amplitude can depend critically on its phase relative to the average, accepted \krzb{} 
phase. In order to maximize our sensitivity
to the non-resonant amplitude, for each phase assumption and \qsq{} bin we made
our measurement based on three \mkpi{} mass regions:
$0.8 < \mkpi < 0.9~\gevcsq $, $0.8 < \mkpi < 1.0~\gevcsq $, and
$0.9 < \mkpi < 1.0~\gevcsq $, which puts the average \krzb{} reference phase at 
roughly $3 \pi/4 $, $3 \pi/2 $, and $7 \pi/4 $ for these three mass regions, respectively.
We chose the mass region with the smallest expected error according to the Monte Carlo 
simulation. Under the assumption
$h^{(d,f)}_0(\qsq{}) = \Hzer{}$, used in Ref.~\cite{formfactor}, 
we performed a $\chi^2$ fit of Fig.~\ref{BESTFD} to the form
$A_{d,f}\,\sin{\delta_{d,f}}\,H^2_0(\qsq)$ over the 
region $\qsq{} < 0.6~\textrm{GeV}^2/c^4$ to
find the amplitude and limits shown in Fig.~\ref{wavelimits}.

Figure~\ref{wavelimits} shows that this ``mass selection'' 
method produced non-amplitude limits, which are reasonably independent of assumed phase. 
If, on the other hand, one used
the full $0.8 < \mkpi < 1.0 ~\gevcsq $ mass range for all sixteen phase assumptions, one would 
get dramatically poorer limits for phase
choices orthogonal to the Breit-Wigner amplitude phase. 
It is apparent from Fig.~\ref{wavelimits} that we have no compelling evidence
for either a $d$-wave, or an $f$-wave component. 
\mysection{\label{sum}SUMMARY}
We present a branching fraction and form factor analysis of the \kpilndk{} decay
based on a sample of approximately 11800 \kpiendk{} and \kpimndk{} decays
collected by the \hbox{CLEO-c} detector running at the $\psi(3770)$.
We find ${\cal B}_{e}(\krzendk{}) = (5.52\pm0.07\pm0.13)\%$ and 
${\cal B}_\mu(\krzmndk)= (5.27\pm0.07\pm0.14)\%$. 
Our direct measurement of the relative semimuonic to semielectric 
branching ratio using  
Eq.~(\ref{eq:BrRatio-yield})
is $\brat{} = (94.64 \pm 1.95 \pm 1.03)\%$. 

We also present a non-parametric analysis of the helicity basis form factors 
that
control the kinematics of the \kpilndk{} decays. We used a projective weighting
technique that allows one to determine the helicity form factor products
by weighted histograms rather than likelihood based fits. 
We find consistency with the spectroscopic pole dominance model for the 
dominant \Hspls{}, \Hsmin{} and \Hszer{} form factors.   Our measurement on the 
\hoHO{} form factor product suggests that the $h_0$ form factor falls faster 
than
$H_0$ with increasing \qsq{}. 
The form factors 
determined using \kpimndk{} decays are consistent with those determined
using \kpiendk{} decays and are consistent with our earlier study \cite{oldff} 
of \kpiendk{}.
Our measured \Hsmin{} form factor data are more consistent with axial and vector form factors
with the expected spectroscopic pole dominance \qsq{} dependence than with constant axial and vector
form factors. 

Our measurements of the \HsT{} and \HTHO{}
form factor suggests a much smaller \HT{} form factor than expected in Lattice Gauge Theory models \cite{lqcd}.
Within the context of the spectroscopic pole dominance model Eq.~(\ref{SPDpole}), our \HTHO{} measurements
are most consistent with a small \HT{} form factor contribution implying a very
negative value for $r_3 \equiv A_3(0)/A_1(0)$, such as $\rthree = -10$, which would place the predicted 
\brat{} relative branching ratio close
to the phase space estimate of 91\%.  
Finally, we have searched for possible $d$-wave or $f$-wave non-resonant
interference contributions to \kpilndk{}. We have no statistically significant  
evidence for $d$-wave or $f$-wave interference, but are only able to limit these terms 
to roughly less than 1.0 and 1.5 times the observed $s$-wave interference for $d$-wave 
and $f$-wave respectively.

\begin{acknowledgments}
We gratefully acknowledge the effort of the CESR staff
in providing us with excellent luminosity and running conditions.
D.~Cronin-Hennessy thanks the A.P.~Sloan Foundation.
This work was supported by
the National Science Foundation,
the U.S. Department of Energy,
the Natural Sciences and Engineering Research Council of Canada, and
the U.K. Science and Technology Facilities Council.
\end{acknowledgments}


\begin{thebibliography}{99}
\bibitem{bigi}
S.~Bianco, F.L.~Fabbri, D.~Benson, and I.~Bigi, Riv. Nuovo Cim. \textbf{26N7},
1 (2003).

\bibitem{quark}
M.~Bauer, B.~Stech, and M.~Wirbel, Z. Phys. C \textbf{29}, 637 (1985);
M.~Bauer and M.~Wirbel, Z.~Phys. C \textbf{42}, 671 (1989);
J.G.~Korner and G.A.~Schuler, Z.~Phys.~C \textbf{46}, 93 (1990);
F.J.~Gilman and R.L.~Singleton, Phys. Rev. D \textbf{41}, 142 (1990);
D.~Scora and N.~Isgur, Phys. Rev. D \textbf{52}, 2783 (1995);
B.~Stech, Z.~Phys. C \textbf{75}, 245 (1997);
D.~Melikhov and B.~Stech, Phys. Rev. D \textbf{62}, 014006 (2000).

\bibitem{ball}
P.~Ball, V.M.~Braun, H.G.~Dosch, and M.~Neubert, Phys. Lett. B \textbf{259},
481 (1991);
P.~Ball, V.M.~Braun, and H.G.~Dosch, Phys. Rev. D \textbf{44}, 3567 (1991).

\bibitem{lqcd}
C.W.~Bernard, A.X.~El-Khadra, and A.~Soni, Phys. Rev. D \textbf{45}, 869
 (1992);
V.~Lubicz, G.~Martinelli, M.S.~McCarthy, and C.T.~Sachrajda, Phys. Lett. B
 \textbf{274}, 415 (1992);
A.~Abada \etal, Nucl. Phys. B \textbf{416}, 675 (1994);
K.C.~Bowler \etal ~(UKQCD Collaboration),  Phys. Rev. D \textbf{51}, 4905 (1995);
T.~Bhattacharya and R.~Gupta, Nucl. Phys. B (Proc. Suppl.) \textbf{47}, 481 
 (1996);
C.R.~Alton \etal ~(APE Collaboration), Phys. Lett. B \textbf{345}, 513 (1995);
S.~Gusken, G.~Siegert, and K.~Schilling, Prog. Theor. Phys. Suppl. \textbf{122},
 129 (1996);
A.~Abada \etal ~(SPQcdR Collaboration), Nucl. Phys. Proc. Supp. \textbf{119}, 625
 (2003).
\bibitem{analyticity}
  C.~Bourrely, B.~Machet and E.~de Rafael, Nucl.\ Phys.\ B {\bf 189}, 157 (1981);
  C.~G.~Boyd, B.~Grinstein and R.~F.~Lebed, Phys.\ Rev.\ Lett.\  {\bf 74}, 4603 (1995);
  L.~Lellouch, Nucl.\ Phys.\ B {\bf 479}, 353 (1996);
  C.~G.~Boyd, B.~Grinstein and R.~F.~Lebed, Nucl.\ Phys.\ B {\bf 461}, 493 (1996);
  I.~Caprini and M.~Neubert, Phys.\ Lett.\ B {\bf 380}, 376 (1996);
  I.~Caprini, L.~Lellouch and M.~Neubert, Nucl.\ Phys.\ B {\bf 530}, 153 (1998);
  C.~G.~Boyd and M.~J.~Savage, Phys.\ Rev.\ D {\bf 56}, 303 (1997);
  M.~Fukunaga and T.~Onogi, Phys.\ Rev.\ D {\bf 71}, 034506 (2005);
  C.~M.~Arnesen, B.~Grinstein, I.~Z.~Rothstein and I.~W.~Stewart, Phys.\ Rev.\ Lett.\ {\bf 95}, 071802 (2005);
  T.~Becher and R.~J.~Hill, Phys.\ Lett.\  B~{\bf 633}, 61 (2006).

\bibitem{slovenia}
S.~Fajfer and J.~Kamenik, Phys. Rev. D~\textbf{72}, 034029 (2005).

\bibitem{helicity-focus}
J.M.~Link \etal ~(FOCUS Collaboration), Phys. Lett. B~\textbf{633}, 183 (2006). 
\bibitem{anomaly}
J.M.~Link \etal ~(FOCUS Collaboration), Phys. Lett. B~\textbf{535}, 43 (2002).
\bibitem{formfactor}
J.M.~Link \etal ~(FOCUS Collaboration), Phys. Lett. B~\textbf{544}, 89 (2002).
\bibitem{KS}
J.G.~Korner and G.A.~Schuler, Z.~Phys.~C~\textbf{46}, 93 (1990);
Fredrick~J.~Gilman
 and Robert~L.~Singleton,~Jr. Phys. Rev. D~\textbf{41}, 142 (1990)
\bibitem{detector}
Y.~Kubota \etal ~(CLEO Collaboration), Nucl. Instrum. Methods A~\textbf{320}, 66 (1992);
M.~Artuso \etal, Nucl. Instrum. Methods A~\textbf{554}, 147 (2005);
D.~Peterson \etal, Nucl. Instrum. Methods A~\textbf{478}, 142 (2002). 
\bibitem{HadronicBrFraction}
S.~Dobbs {\it et al.} (CLEO Collaboration), Phys. Rev. D~{\bf 76}, 112001 (2007).
\bibitem{evtgen}
D.J.~Lange, Nucl. Instrum. Methods  A~\textbf{462}, 152
(2001).
\bibitem{geant}
R.~Brun \etal, {\bf Geant 3.21}, CERN Program Library Long Writeup W5013,
unpublished.
\bibitem{absBeCLEO}
G.S.~Hung {\it et al.} (CLEO Collaboration) Phys. Rev. Lett.~\textbf{95}, 181801 (2005).
\bibitem{pdg}
C.~Amsler et al. (Particle Data Group), Phys. Lett. B~\textbf{667}, 1 (2008).
\bibitem{oldff}
M.R.~Shepherd \etal ~(CLEO Collaboration), Phys. Rev. D~\textbf{74}, (2006) 052001. 
\end{thebibliography}
\end{document}